\documentclass[journal,a4paper]{IEEEtran}
\normalsize
\usepackage[english]{babel}
\usepackage{ragged2e}
\usepackage{blindtext}
\usepackage{graphicx}
\usepackage{epstopdf}
\usepackage{multicol}
\usepackage{diagbox}
\usepackage{blkarray}
\usepackage{romannum}

\usepackage{hyperref}
\usepackage{float}
\usepackage{url}
\usepackage[utf8]{inputenc} 
\usepackage{array}
\usepackage[utf8]{inputenc} 
\usepackage[T1]{fontenc}
\usepackage{amsmath}
\usepackage{mathtools}
\usepackage{amsfonts}
\usepackage{amssymb}
\usepackage{amsthm}
\usepackage{enumerate}
\usepackage{cite}
\usepackage{calc}
\usepackage{color}
\usepackage{epsfig}
\usepackage{setspace}
\usepackage{pstricks}
\usepackage{cancel}
\usepackage{multirow}
\usepackage{mathrsfs}
\usepackage{algorithm}
\usepackage{algpseudocode}
\usepackage{multirow}
\usepackage{romannum}
\usepackage{diagbox}
\usepackage{booktabs,makecell}
\usepackage{mathrsfs}
\usepackage{float}
\usepackage{enumitem}

\theoremstyle{definition}
\newtheorem{example}{Example}
\graphicspath{{./images/}}
\def\Ddots{\mathinner{\mkern1mu\raise\p@
		\vbox{\kern7\p@\hbox{.}}\mkern2mu
		\raise4\p@\hbox{.}\mkern2mu\raise7\p@\hbox{.}\mkern1mu}}
\makeatother

\begin{document} 
	\pagenumbering{arabic}
	\title{Design and Analysis of Index Codes for 3-Group NOMA in Vehicular Adhoc Networks}
	\author{
		\IEEEauthorblockN{Sai Pavan Deekshitula and B. Sundar Rajan, \textit{Fellow, IEEE}}\\
		\IEEEauthorblockA{Department of Electrical Communication Engineering, Indian Institute of Science, Bengaluru.\\
			E-mail: saipavand@iisc.ac.in, bsrajan@iisc.ac.in.}
	}
	\maketitle

	\begin{abstract} 
	Index coding (IC) is a source coding technique employed to improve spectral utilisation, where the source node aims to satisfy users' demands by making minimum transmissions. 
	Non-orthogonal multiple access (NOMA) is integral to the radio access technique used in 5G networks. Index-coded NOMA (IC-NOMA) transmission scheme in Vehicular Adhoc Networks (VANETs) involves applying NOMA principles on index-coded data to avoid network congestion and to improve spectral efficiency compared to conventional IC systems. In this work, a spectral efficient transmission scheme called 3-Group IC-NOMA is proposed, and an innovative index code design that fits with NOMA decoding principles to obtain improved spectral efficiency is developed. Through exhaustive analytical studies, we demonstrate that the proposed transmission scheme always supports higher rates than the conventional IC systems and requires less power to achieve an information rate at least as good as conventional IC systems. {\footnote{Part of the manuscript has been communicated to IEEE International Symposium on Personal, Indoor and Mobile Radio Communications (PIMRC 2023) to be held in September 2023.}}
	
\end{abstract}


\section{Introduction}
\label{intro}
Vehicular Adhoc Networks (VANETs) enable vehicle-to-everything (V2X) communication and have become an important area of research, standardisation and development in recent years. Applications in VANETs may be classified into road safety, traffic efficiency and infotainment types \cite{CHS}. The performance requirements of different VANET applications differ from each other. For example, road safety applications require low latency and high reliability. In contrast, infotainment applications have high bandwidth and QoS-sensitive requirements.

The choice of source coding technique employed in VANETs is essential as we have bandwidth limitations and require fast data distribution \cite{ISE}. Birk and Kol introduced a source coding variant called Index Coding (IC) in \cite{ZYT}. The participating users are a set of caching clients. These users possess some prior information in their cache called as side information. Side information can be a set of message packets or a linear combination of message packets, called as coded side information\cite{NAR} \cite{KMC}. Through a backward channel, these clients inform the central server about the side information they possess in their respective caches. The central server has to use the knowledge of each client's side information and find a way to satisfy every client's request using minimum transmissions. Employing the IC technique in VANETs can increase overall transmission efficiency as it reduces the network load by meeting every client demand by making minimum transmissions.

Non-orthogonal multiple access (NOMA) is a promising radio access technique in next-generation wireless communications. NOMA offers enhanced spectrum efficiency \cite{PJL}, reduced latency with high reliability, and accommodates more users compared to orthogonal frequency division multiple access (OFDMA), the current de facto standard of orthogonal multiple access (OMA) techniques. On the transmitter side, if individual waveforms of different users are combined in the power domain, it is called power domain NOMA (PD-NOMA)\cite{SNO}. Else, if the individual waveforms are combined in the code domain, it is called code domain NOMA (CD-NOMA). Power domain multiplexing \cite{LBY} \cite{ZLG}  at the transmitter and successive interference cancellation (SIC) at the receiver, enables NOMA to serve various users using the same resources in terms of time, frequency and space. Employing NOMA in VANETs can increase the overall spectral efficiency. This work concerns only PD-NOMA; hence, in all further discussions, the term NOMA is used instead of PD-NOMA.

Authors in \cite{PPM} have proposed a transmission strategy in VANETs called Index Coded-NOMA (IC-NOMA) by integrating IC concepts with NOMA. They have considered a VANET scenario where $N$ vehicles move at almost the same speed on a multilane track. Each vehicle is equipped with an LTE-V communication device, and there is no cooperative communication between these vehicles due to privacy and security issues. The users were classified into far and near groups depending on channel conditions. To reap the benefits of NOMA, they have proposed an index code design wherein the index codes for far and near users were developed separately. The algorithm proposed in \cite{PPM} to design index codes for far and near users can be summarised as follows: An index code for far group users is designed considering their side information only. In NOMA decoding, near users can obtain far users' data for SIC. So, the index code for near users is developed by considering the near user's side information, and the index code designed for far users as additional coded side information. Then the far and near users' index codes are suitably combined in the power domain to perform NOMA transmissions. In this work we consider a more practical scenario and extend the grouping of users into three groups- far users, intermediate users and near users- and show that more complicated and interesting scenario arise.

	The main technical contributions of this paper may be summarized as follows:  
\begin{itemize}
	\item We introduce a spectral efficient transmission scheme called the “3-Group IC-NOMA” scheme for VANETS involving three groups of users and propose an innovative index code design for the same which constitutes a non-trivial extension of \cite{PPM}, allowing us to reap the benefits of both IC and NOMA.  
	\item We demonstrate that our transmission scheme serves user demands in less than or equal number of transmissions than the conventional IC systems, leading to increased spectral efficiency.  
	\item Through detailed analytical studies, we show that our transmission scheme always supports a higher sum rate of the transmission  and provides positive power savings compared to conventional IC systems to achieve a sum rate of at least as good as conventional IC systems.
\end{itemize}  

\section{Preliminaries}
\label{Prelim}
\subsection{Index coding}
\label{Indexcoding}
In an index coding problem (ICP), a single sender, i.e. a central server, desires to transmit messages to multiple receivers over a broadcast channel\cite{SVY}. The server has a set of $n$ messages represented as \(\mathcal{X =} \) \{$x_{1},x_{2}...x_{n}\}$. Each $x_i$ $\in$ $F_{q}$,  where $F_{q}$  denotes a finite field of size \textit{q}. There are $N$ receivers, and each receiver is denoted as \textit{$V_i$}, $i$ $\in$ [$N$], where [$N$] denotes set of first N natural numbers. Each receiver \textit{$V_i$} needs a subset of these messages known as want set, denoted by \(\mathcal{W} \)$_i$ $\subseteq$ \(\mathcal{X } \) and already holds a subset of these messages termed as known set, denoted by \(\mathcal{K} \)$_i$, with \(\mathcal{K} \)$_i$  $\subseteq$  \(\mathcal{X} \). The known set of the receiver is called side information possessed by that receiver. Let the known set and want set of the ICP be denoted as \(\mathcal{K}\) = \{\(\mathcal{K} \)$_i$: \textit{i} $\in$ [$N$]\} and  \(\mathcal{W}\) = \{\(\mathcal{W} \)$_i$: \textit{i} $\in$ [$N$]\}. An ICP with \textit{n} messages, \textit{N} receivers, known set 
\(\mathcal{K}\) and want set \(\mathcal{W}\)  is denoted as \(\mathcal{I} \)(\textit{n}, \textit{N}, \(\mathcal{K} \), \(\mathcal{W} \)).

\textbf{\textit{Definition 1}}: An index code \textbf{y} $\in$ $F_{q}^l$ for an instance of  \(\mathcal{I} \)(\textit{n}, \textit{N}, \(\mathcal{K} \), \(\mathcal{W} \)) ICP  with input vector \textbf{x} $\in$ $F_{q}^n$, consists of an encoding function \(\mathcal{F} \) : $F_{q}^n$ $\rightarrow$ $F_{q}^l$ and a corresponding decoding function for each receiver  \(\mathcal{G}\)$_i$ : $F_{q}^l$ $\times$  $F_{q}^{|\textit{K}_i|}$ $\rightarrow$ $F_{q}^{|\textit{W}_i|}$;  \textit{i} $\in$ [\textit{N}] and $l$ is called the length of the index code. If $l$ is minimum, then the index code is considered optimal. 

If the encoding function \(\mathcal{F} \) is linear over $F_{q}$, then the index code is said to be \textit{linear}. A linear index code \textbf{y} can be written as $\textbf{y}$$=$$\textbf{L}$$\textbf{x}$ where \textbf{L} is a \textit{l} $\times$ \textit{n} matrix over $F_{q}$ known as \textit{encoding matrix}.

Authors in \cite{NAR} and \cite{KMC}  generalised the IC problem where the elements of the known set are linear combinations of messages. Such IC problems are known as index coding with coded side information(ICCSI). The index coding vector for the ICCSI problem is given as $\textbf{y}_c$=$\textbf{L}_c$\textbf{x}, where $\textbf{L}_c$ is the encoding matrix used at the sender.

	\begin{example}
	\label{ex:1}
	The server has a set of 4 messages represented as \(\mathcal{X =} \) \{$x_{1},x_{2},x_{3},x_{4}\}$. Let the known set and want set of the ICP with $\textit{N=}4$ users be \(\mathcal{K}\) = \{$x_i$: \textit{i} $\in$ [4]\} and \(\mathcal{W}\) = \{$x_{2},\{x_{1},x_{3}\},\{x_{2},x_{4}\},x_{3}\}$. 
	
	The  index-coded transmissions are given by $y_1$ = $x_1$ + $x_2$, $y_2$ = $x_2$ + $x_3$ and $y_3$ = $x_3$ + $x_4$. The length of this index code is 3. The decoding process at  users $V_1,V_2,V_3$ and  $V_4$ respectively  is given by  $y_1$ +  $x_1$ = $x_2$, \{$y_1$ +  $x_2$ = $x_1$, $y_2$ +  $x_2$ = $x_3$\}, \{$y_2$ +  $x_3$ = $x_2$, $y_3$ +  $x_3$ = $x_4$\} and $y_3$ +  $x_4$ = $x_3$.
\end{example}

\subsection{Non-orthogonal multiple access}
\label{NOMA}
In NOMA, signals of multiple users are linearly combined in the power domain depending on their channel conditions. Users that are served simultaneously by a single superposed transmission form a cluster.\cite{MTE}

	Assuming there are \textit{M} users per cluster, each user is denoted as $V_i$ with $i$ $\in$ [$M$]. Let $x_i$ be the data requested by user $V_i$,  $V_M$ be the farthest user, $V_1$ be the nearest user and the channel gains be $g_1$ > $g_2$ >......> $g_M$. Let $s_i$ be the encoded form of $x_i$ i.e $s_i$$=$enc($x_i$). Let the additive white Gaussian noise with zero mean and unit variance at receiver $V_i$ be denoted as $n_i$.

Considering the channel conditions of each user, signals requested by all users are combined linearly in the power domain at the transmitter. All the users in a cluster are served simultaneously by single superposed transmission $S=\sum_{i=1}^{M} \sqrt{\alpha_i P}s_i$,
where $P$ is the transmission power, $\alpha_i$$P$ is the amount of power allocated to signal $s_{i}$ corresponding to user $V_i$ with $\sum_{i=1}^{M} $$\alpha_i$ = 1 and $\alpha_1$ < $\alpha_2$ < $\alpha_3$ < \dots < $\alpha_M$.
The received signal $Z_i$ at $V_i$ is given by $Z_i=\sqrt{g_i}S + n_i$.
The far user $V_M$  decodes the desired signal by considering interference from all the other users as noise. All other users perform SIC\cite{XHW}, where the highest power data is decoded and progressively cancelled to decode the desired signal.

Consider a NOMA system with $M$=3. In this case, the transmitted signal is given by 
$S$$=$$\sqrt{\alpha P}$$S_1$ + $\sqrt{\beta P}$$S_2$ + $\sqrt{\gamma P}$$S_3$;  where $\alpha$ < $\beta$ < $\gamma$ and $\alpha$ + $\beta$ + $\gamma$$=$1. Here $S_3$$=$enc(\textit{$x_{3}$}) is the signal intended for far user $V_3$, $S_2$$=$enc(\textit{$x_{2}$}) is the signal intended for intermediate user $V_2$ and $S_1$$=$enc(\textit{$x_{1}$}) is the signal intended for near user $V_1$. Users $V_1$, $V_2$ and $V_3$ receive $Z_1$$=$$\sqrt{g_1}$$S$ + $n_1$, $Z_2$$=$$\sqrt{g_2}$$S$ + $n_2$ and $Z_3$$=$$\sqrt{g_3}$$S$ + $n_3$ respectively. Far user $V_3$ decodes $x_3$ by considering interference due to $x_1$ and $x_2$ as noise. Intermediate user $V_2$ performs SIC of $x_3$. After SIC, the signal at $V_2$ can be represented as $Z_2^{'}$$=$$Z_2$$-\sqrt{g_2 \gamma P}$$S_3^{'}$ with $S_3^{'}$ as signal intended for $V_3$ decoded at $V_2$. Intermediate user $V_2$ decodes for  $x_{2}$ from $Z_2^{'}$.  Similarly $V_1$ performs SIC of $x_2$ and $x_3$. After SIC the signal at $V_1$ can be represented as $Z_1^{'}$$=$$Z_1$$-\sqrt{g_1 \gamma P}$$S_3^{''}$$-\sqrt{g_1 \beta  P}$$S_2^{''}$
with $S_2^{''}$ and $S_3^{''}$ as signals intended for $V_2$ and $V_3$ decoded at $V_1$.  Near user $V_1$ decodes for  $x_{1}$ from $Z_1^{'}$.

%

\section{Index Code Design for 3-Group IC-NOMA system}
\label{Design}

\subsection{System Model}
	We consider a downlink VANET scenario \cite{PPM} in which vehicles travel on a one-way multilane road at speeds such that their relative velocity is negligible. These vehicles could be interested in common information\cite{WDW}, while each can have distinct demands for entertainment data. Each vehicle is equipped with an LTE-V-based communication device\cite{SJY}, and there is no cooperative communication\cite{QMZ} between vehicles due to security and privacy issues. This work considers LTE-V or Cellular-V2X standard (C-V2X) \cite{SJYL}, and thus, the vehicles can exchange information with infrastructures such as roadside units (RSU) or base stations (BS). Each vehicle is equipped with a cache where the data packets delivered by RSU or BS are stored as side information. The scenario under consideration consists of 3 different phases:
\newline
1)\textbf{Reporting Phase}: The vehicles report their demands to the server through RSU or BS. The demand set of vehicles can have some packets demanded in common. Upon receiving user demands, the server sends the total demand set to the next RSU through BS in the forward direction of the vehicles through wireline channels.
\newline
2)\textbf{R2V Phase}: The RSU disseminates data packets to users in its transmission range. Each user receives a subset of packets delivered by RSU as their side information.
Given the range of RSU, the speed of the vehicles, and the number of messages demanded, it may or may not satisfy their demands completely. 
\newline
3)\textbf{V2N Phase}: The vehicles communicate only with BS in this phase. The set of vehicles entering the range of BS form a cluster. The vehicles will send the index set of their side information through a wireless back channel to BS. The BS updates the demand set of vehicles by removing the messages they received in the R2V phase.
\vspace{-0.2cm}

\subsection{3-Group IC-NOMA in VANET} 
	Consider a cluster of $N$ vehicles with \(\mathcal{D}\)$_i$, $i$ $\in$ [$N$] as the initial demand set of vehicle $V_i$, which it communicates to the server via RSU or BS in reporting phase. The server through BS sends the whole set of demanded packets denoted as \(\mathcal{X}\) = $\cup_{i\in[N]}$\(\mathcal{D}\)$_i$ to the next RSU in the direction of vehicle movement.

During the R2V phase, RSU disseminates data packets to users in its transmission range, and user $V_i$, $i$ $\in$ [$N$] will receive the side information packets denoted as \(\mathcal{K}\)$_i$.  In the V2N phase, BS updates the demand set of each user by removing the messages they received as side information in the R2V phase. Let the updated demand set of $V_i$ be denoted as
\(\mathcal{W}\)$_i$$=$\(\mathcal{D}\)$_i$$\setminus$\{\(\mathcal{D}\)$_i$ $\cap$ \(\mathcal{K}\)$_i$\}.

To apply NOMA principles to index coded data, the vehicles in a cluster are further divided into far, intermediate and near groups. Let $N$$_f$, $N$$_m$ and $N$$_n$ denote the number of users in far, intermediate and near groups, respectively. Let $g_i$, $i$ $\in$ [$N$] denote channel gain between vehicle $V_i$ and BS. Without loss of generality, let $g_1$$\ge$$g_2$$\ge$$\dots$$\ge$$g_{N_{n}}$$\ge$$\dots$$\ge$$g_{N_{n}+N_{m}}$$\dots$$\ge$$g_N$. 

\vspace{0.1cm}
Let $l^{IC-NOMA} $ denote the minimum number of transmissions required by the proposed transmission scheme.
The $j^{th}$  signal received by $V_i$ can be represented as.            
\begin{equation*}            	
	\begin{split}
		Z_{i}^{j} = \sqrt{g_i}S_{j} + n_{i} ,i  \in  [N], j  \in  \left [l^{IC-NOMA} \right ]
	\end{split}
\end{equation*}
where $S_{j}$ represents the $j^{th}$ transmitted signal, $g_i$ reprsents the channel gain between the BS and user $V_i$ and $n_i$ reprsents additive white Gaussian noise of zero mean and variance $\sigma_{i}^{2}$ at user $V_i$. We call this a 3-Group IC-NOMA system, as the vehicles in the cluster are divided into three groups.

\begin{figure}
	\centering
	\includegraphics[width= \columnwidth]{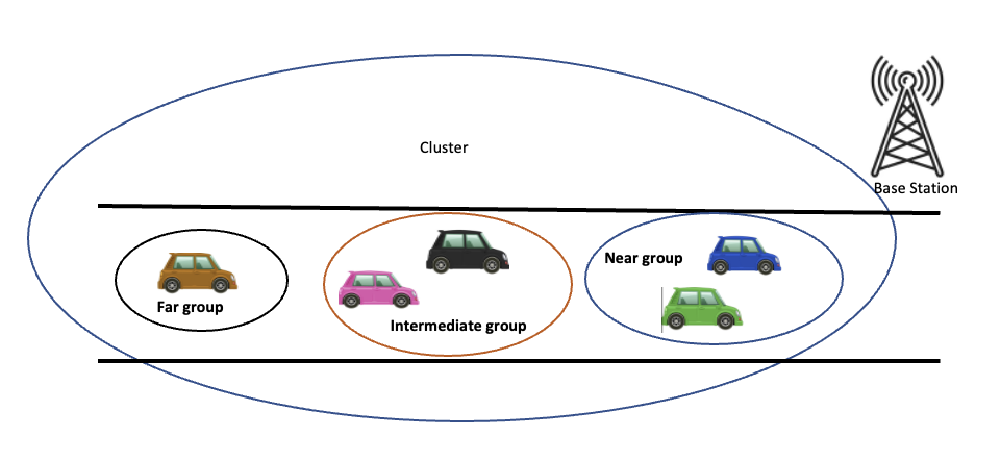}
	\caption{Vehicles in the cluster divided into three Groups based on channel conditions}
	\label{SM}
\end{figure}

\subsection{Motivating Example} 
\label{ME}
	\begin{example}
	\label{ex-2}
	The server has a set of 4 messages represented as \(\mathcal{X =} \) \{$x_{1},x_{2},x_{3},x_{4}\}$. Let the known set and want set of the ICP with $\textit{N=}4$ users be  \(\mathcal{K}\) = \{$x_{2}, x_{1}, \phi, \phi\}$ and \(\mathcal{W}\) = \{$x_i$: \textit{i} $\in$ [4]\}.
	
	The  index coded transmissions are $y_1$ = $x_1$+$x_2$, $y_2$ = $x_3$ and $y_3$ = $x_4$. Hence, in a conventional IC scenario, three transmissions are required to satisfy the demands of users.
	
	In this example, let $V_1$, $V_2$ be near users, $V_3$ be the intermediate user and $V_4$ be the far user so that $N_n$ = 2, $N_m$ = 1 and $N_f$ = 1. The 3-Group IC-NOMA transmissions combine the concepts of NOMA and IC to transmit the signals $s_1$ = enc($y_1$), $s_2$ = enc($y_2$) and $s_3$ = enc($y_3$).
	
	The transmitted signal in this case is $S$= $\sqrt{\alpha P}$$s_1$$+\sqrt{\beta P}$$s_2$$+\sqrt{\gamma P}$$s_3$ with  $\alpha$ < $\beta$ < $\gamma$ and $\alpha$ + $\beta$ + $\gamma$ = 1. Here, data requested by two near users are paired by index coding under the same power level. The far user, $V_4$, gets $x_4$ transmitted at the highest power level. The intermediate user, $V_3$, decodes $y_3$ = $x_4$ and performs SIC to get index-coded packet $x_3$. The near users decode $y_3$ = $x_4$ and $y_2$ = $x_3$ and perform SIC to get index coded packet $y_1$ = $x_1$+$x_2$. They exploit their side information to meet their demands. Hence, the required number of 3-Group IC-NOMA transmissions in  {\textit{Example}}-\ref{ex-2} is given as $l^{IC-NOMA}$$=$1.
	
\end{example}     					

\subsection{Algorithm for Index Code Design for 3-Group IC-NOMA System}  
\label{ssec-1}

\begin{algorithm}
	\caption{Grouping of users and Design of Index codes in 3-Group IC-NOMA system}
	\label{alg:1}
	\begin{algorithmic}[1]
		\Require $g_i$, \(\mathcal{K}\)$_i$ and \(\mathcal{W}\)$_i$ $\forall$ $i$ $\in$ [$N$], $x$=[$x_1$,$x_2$,...$x_n$]$^{T}$, $N$
		
		\State $g_{max}$$=$$max_{i \in [N]}(g_i), g_{med}$$=$$median_{i \in [N]}(g_i)$\, $g_{min}$$=$$min_{i \in [N]}(g_i).$
		
		\State \(\mathcal{I}\)$_f$ = [$N$] 
		, \(\mathcal{I}\)$_m$ = [$N$]
		, \(\mathcal{I}\)$_n$ = [$N$].
		
		\For{i = 1 to $N$}  
		\If{|$g_{max}$ - $g_i$| < $min$(|$g_{med}$ - $g_i$|,|$g_{min}$ - $g_i$|)}
		
		\State \(\mathcal{I}\)$_m$ = \(\mathcal{I}\)$_m$ $\setminus$ \{$i$\} 
		\State \(\mathcal{I}\)$_f$  = \(\mathcal{I}\)$_f$ $\setminus$ \{$i$\}

		\ElsIf{|$g_{med}$ - $g_i$| < $min$(|$g_{max}$ - $g_i$|,|$g_{min}$ - $g_i$|)}
		\State \(\mathcal{I}\)$_n$ = \(\mathcal{I}\)$_n$ $\setminus$ \{$i$\} 
		\State \(\mathcal{I}\)$_f$ = \(\mathcal{I}\)$_f$ $\setminus$ \{$i$\} 
		
		\Else
		\State \(\mathcal{I}\)$_m$ = \(\mathcal{I}\)$_m$ $\setminus$ \{$i$\} 
		\State \(\mathcal{I}\)$_n$  = \(\mathcal{I}\)$_n$ $\setminus$ \{$i$\} 
		\EndIf
		
		\EndFor

		\State  \(\mathcal{K}\)$_f$ = {\{\(\mathcal{K}\)$_i$\ : $\textit{i}$ $\in$ \(\mathcal{I}\)$_f$\}} ,
		\(\mathcal{W}\)$_f$\ = {\{\(\mathcal{W}\)$_i$\ : $\textit{i}$ $\in$  \(\mathcal{I}\)$_f$\}} .
		
		\State Design encoding matrix $\textbf{L}$$_f$.
		
		\State $\textbf{y}$$_f$ = $\textbf{L}$$_f$$\textbf{x}$.
		\State Let \(\mathcal{Y}_f \) $\equiv$ {\{$y$$_f$, $\textit{i}$ $\in$ [$\textit{l}$$_f$]\}}  denote the set of index coded packets designed 
		for far users.
		\vspace{0.1cm}									
		\State \(\mathcal{K}\)$_i^{'}$\ $=${\{\(\mathcal{K}\)$_i$ $\cup$ \(\mathcal{Y}_f \) : $\textit{i}$ $\in$ \(\mathcal{I}\)$_m$\}}
		\vspace{0.1cm}
		\State \(\mathcal{K}_m^{'} \)$=${\{\(\mathcal{K}\)$_i^{'}$\ : $\textit{i}$ $\in$ \(\mathcal{I}\)$_m$\}} 
		
		\State Let {\{\(\mathcal{W}_i^{c} \) : $\textit{i}$ $\in$ \(\mathcal{I}\)$_m$\}} denote the set of demands of intermediate user  $V_i$ 
		satisfied by the index code of far users.
		
		\State \(\mathcal{W}\)$_m^{'}$\ $\equiv${\{\(\mathcal{W}\)$_i$ $\setminus$ \(\mathcal{W}_i^{c} \) : $\textit{i}$ $\in$ \(\mathcal{I}\)$_m$\}}
		\State Design encoding matrix $\textbf{L}$$_m^{c}$
		\State $\textbf{y}$$_m^{c}$$=$$\textbf{L}$$_m^{c}$$\textbf{x}$.
		\State Let \(\mathcal{Y}_m\)$\equiv${\{$y$$_m^{c}$, $\textit{i}$ $\in$ [$\textit{l}$$_m$]\}} denote the set of index coded packets designed 
		for intermediate users.
		
		\State \(\mathcal{K}\)$_i^{''}$\ $=${\{\(\mathcal{K}\)$_i$ $\cup$ \(\mathcal{Y}_f \) $\cup$ \(\mathcal{Y}_m \) : $\textit{i}$ $\in$ \(\mathcal{I}\)$_n$\}}
		
		\State \(\mathcal{K}_n^{'} \)={\{\(\mathcal{K}\)$_i^{''}$\ : $\textit{i}$ $\in$ \(\mathcal{I}\)$_n$\}} 
		\State Let {\{\(\mathcal{W}_i^{c} \) : $\textit{i}$ $\in$ \(\mathcal{I}\)$_n$\}} denote the set of demands of near user  $V_i$ 
		satisfied by coded side information of far users and 
		intermediate users.
		
		\State\(\mathcal{W}\)$_n^{'}$\ $\equiv${\{\(\mathcal{W}\)$_i$ $\setminus$ \(\mathcal{W}_i^{c} \) : $\textit{i}$ $\in$ \(\mathcal{I}\)$_n$\}}
		\State Design encoding matrix $\textbf{L}$$_n^{c}$.
		
		\State $\textbf{y}$$_n^{c}$ = $\textbf{L}$$_n^{c}$$\textbf{x}$.
		\State \textbf{return}  $\textbf{y}$$_f$ , $\textbf{y}$$_m^{c}$ , $\textbf{y}$$_n^{c}$.

	\end{algorithmic}
\end{algorithm}

In a conventional IC scenario, an optimal index code can be designed based on the want set and the known set of all users. In the 3-Group IC-NOMA transmission scheme,
near users apply SIC for NOMA decoding and can retrieve information intended for intermediate and far group users. So, index-coded data sent to the far and intermediate group users can be used as additional coded side information to meet the demands of near users. Also, the intermediate users apply SIC for decoding and can retrieve information intended for far group users. So, the index-coded data sent to the far users can be used as additional coded side information to meet the demands of the intermediate users. However, the index-coded packets sent at higher power levels must provide a complete IC solution to far users because the packets sent at other power levels are considered noise and discarded. Hence, in the 3-Group IC-NOMA system, index codes for far, intermediate and near group users are designed separately to reap the benefits of NOMA. 

Let $x$$=$[$x_1$,$x_2$,...$x_n$]$^{T}$ be the set of messages to be communicated by the BS after the V2N phase. Let  \(\mathcal{I}\)$_f$\ denote the index set of far users, \(\mathcal{I}\)$_m$\ denote the index set of intermediate users and \(\mathcal{I}\)$_n$\ denote the index set of near users.

Considering   the want set and known set of far users denoted as \(\mathcal{W}\)$_f$\ ={\{\(\mathcal{W}\)$_i$, \ $\textit{i}$ $\in$ \(\mathcal{I}\)$_f$\}} and 
\(\mathcal{K}\)$_f$\ ={\{\(\mathcal{K}\)$_i$,\  $\textit{i}$ $\in$ \(\mathcal{I}\)$_f$\}}, a linear IC problem is formulated.
A linear index code for far users specified by encoding matrix $\textbf{L}$$_f$ is given by
\begin{equation}
	\label{eqn:far}
	\textbf{y}_f = \textbf{L}_f\textbf{x}
\end{equation}
where $\textbf{L}$$_f$ is a $\textit{l}$$_f$$\times$n matrix. The length of the index code designed for far users is denoted as $\textit{l}$$_f$. Let \(\mathcal{Y}_f \)$=${\{$y$$_f$, $\textit{i}$ $\in$ [$\textit{l}$$_f$]\}}  be the set of index coded packets designed for far users.

The intermediate users decode the data of far users to perform SIC. Hence, an index code is designed for intermediate users by considering their respective side information and far users' index code as additional coded side information. The side information at each intermediate user can be modified as \(\mathcal{K}\)$_i^{'}$\  $=$ {\{ \(\mathcal{K}\)$_i$ $\cup$ {\(\mathcal{Y}_f \): $\textit{i}$ $\in$ \(\mathcal{I}\)$_m$\}} and the total known set of intermediate users is represented as  \(\mathcal{K}_m^{'} \) $=$ {\{\(\mathcal{K}\)$_i^{'}$,\ $\textit{i}$ $\in$ \(\mathcal{I}\)$_m$\}}. The total want set of intermediate users is represented as  \(\mathcal{W}\)$_m^{'}$\ = {\{\(\mathcal{W}\)$_i$$\setminus${\(\mathcal{W}_i^{c} \) : $\textit{i}$ $\in$ \(\mathcal{I}\)$_m$\}}, where \(\mathcal{W}_i^{c} \) is the demand set of intermediate user $V_i$ satisfied through index code of far users. Hence, the index coding problem for intermediate group users is an ICCSI problem with encoding matrix $\textbf{L}$$_m^{c}$. The index-coded vector for intermediate users is obtained as 
		\begin{equation}
			\label{eqn:inter}
			\textbf{y}_m^{c} = \textbf{L}_m^{c}\textbf{x}
		\end{equation}
where $\textbf{L}$$_m^{c}$ is  $\textit{l}$$_m$$\times$n matrix. The length of the index code designed for intermediate users is denoted as $\textit{l}$$_m$.    			
		Let  \(\mathcal{Y}_m\)$=${\{$y$$_m^{c}$, $\textit{i}$ $\in$ [$\textit{l}$$_m$]\}} be the set of index coded packets designed for intermediate users.
		
									\begin{algorithm}
	\caption{Transmissions in 3-Group IC-NOMA scheme}
	\label{alg:2}
	\begin{algorithmic}[1]
		\Require  $s_{f_{i}}$, $\forall$ $\textit{i}$ $\in$ [$\textit{l}$$_f]$; $s_{m_{i}}$, $\forall$ $\textit{i}$ $\in$ [$\textit{l}$$_m]$; $s_{n_{i}}$, $\forall$ $\textit{i}$ $\in$ [$\textit{l}$$_n]$; \hspace{1.5cm}$P$ , $\alpha$, $\beta$, $\gamma$, $\alpha_{1}$ such that $\alpha$<$\beta$<$\gamma$ , $\alpha$+$\beta$+$\gamma$$=$1 and $\alpha_{1}$$<$0.5.
		\State \textbf{Initialize} $L$$=$0.
		\State $K$$=$$min$($\textit{l}$$_f$, $\textit{l}$$_m$, $\textit{l}$$_n$).
		\If{K $\neq$ 0}
		\For{\textit{k} = 1 to $K$  }   
		\vspace{0.1cm}
		\State $S_{k}^{NOMA^{(3)}}$ $=$  $\sqrt{\alpha P}s_{n_{k}}+\sqrt{\beta P}s_{m_{k}}+\sqrt{\gamma P}s_{f_{k}}$;
		\EndFor
		\EndIf
		\If{$min$($l_f$,$l_n$) $>$ $l_m$}
		\State $L$$=$$min$($l_f$,$l_n$) $-$ $l_m$
		\For {\textit{k} = $K+1$  to $K+L$}
		\vspace{0.1cm}
		\State $S_{k}^{NOMA^{(2)}}$ $=$	$\sqrt{\alpha_1 P}s_{n_{k}}$+$\sqrt{(1-\alpha_{1}) P}s_{f_{k}}$;
		\EndFor
		\ElsIf{$min$($l_m$,$l_n$) $>$ $l_f$}
		\State $L$$=$$min$($l_m$,$l_n$) $-$ $l_f$
		\For {\textit{k} = $K+1$  to $K+L$}
		\vspace{0.1cm}
		\State $S_{k}^{NOMA^{(2)}}$ $=$	$\sqrt{\alpha_1 P}s_{n_{k}}$+$\sqrt{(1-\alpha_{1}) P}s_{m_{k}}$;
		\EndFor
		\ElsIf{$min$($l_f$,$l_m$) $>$ $l_n$}
		\State $L$$=$$min$($l_f$,$l_m$) $-$ $l_n$
		\For {\textit{k} = $K+1$  to $K+L$}
		\vspace{0.1cm}
		\State $S_{k}^{NOMA^{(2)}}$ $=$	$\sqrt{\alpha_1 P}s_{m_{k}}$+$\sqrt{(1-\alpha_{1}) P}s_{f_{k}}$;
		\EndFor
		\EndIf
		\If{$l_f$ $>$ $max$($l_m$,$l_n$)}
		\State $M$$=$$l_f$$-$$max$($l_m$,$l_n$)
		\For {\textit{k} =  $K+L+1$ to $K+L+M$ }
		\vspace{0.1cm}
		\State $S_{k}^{IC}$ $=$	$\sqrt{P}s_{f_{k}}$;
		\vspace{0.1cm}
		\EndFor
		\ElsIf{$l_m$ $>$ $max$($l_f$,$l_n$)}
		\State $M$$=$$l_m$$-$$max$($l_f$,$l_n$)
		\For {\textit{k} =  $K+L+1$ to $K+L+M$ }
		\vspace{0.1cm}
		\State $S_{k}^{IC}$ $=$	$\sqrt{P}s_{m_{k}}$;
		\vspace{0.1cm}
		\EndFor
		\ElsIf{$l_n$ $>$ $max$($l_f$,$l_m$)}
		\State $M$$=$$l_n$$-$$max$($l_f$,$l_m$)
		\For {\textit{k} =  $K+L+1$ to $K+L+M$ }
		\vspace{0.1cm}
		\State $S_{k}^{IC}$ $=$	$\sqrt{P}s_{n_{k}}$;
		\EndFor
		\EndIf
	\end{algorithmic}
\end{algorithm}
		
								\begin{table*}[htbp]
			\centering   				
			\caption{Cases in 3-Group IC-NOMA transmission scheme}
			\label{Tab:2}
			
			\begin{tabular}{|p{4.05cm} |p{2.2cm}|p{2.2cm}|p{2.2cm}|p{2.2cm}|}
				\hline
				
				$\textbf{Case}$   
				&$l^{NOMA^{(3)}}$& $l^{NOMA^{(2)}}$ &$l^{IN-IC}$&$l^{IC-NOMA}$\\ [1ex] 
				
				\hline
				$\textbf{Case-I}$:        $l_f$ = $l_m$ = $l_n$
				&\centering $l_f$& \centering 0 &\centering 0 &$l_f$\\ [1ex]
				
				\hline
				$\textbf{Case-II}$:       $l_f$ > $l_m$ > $l_n$
				&\centering $l_n$& \centering $l_m$ - $l_n$ &\centering $l_f$ - $l_m$ &$l_f$\\ [1ex]
				
				\hline
				$\textbf{Case-III}$:        $l_n$ > $l_f$ > $l_m$
				&\centering $l_m$& \centering $l_f$ - $l_m$ &\centering $l_n$ - $l_f$ &$l_n$\\ [1ex]
				
				\hline
				$\textbf{Case-IV}$:     $l_m$ > $l_n$ > $l_f$
				&\centering $l_f$& \centering $l_n$ - $l_f$ &\centering $l_m$ - $l_n$ &$l_m$\\ [1ex]
				\hline

				$\textbf{Case-V}$:        $l_f$ > $l_m$ = $l_n$
				&\centering $l_n$& \centering 0 &\centering $l_f$ - $l_m$ &$l_f$\\ [1ex]
				
				\hline
				$\textbf{Case-VI}$:        $l_f$ > $l_n$ > $l_m$
				&\centering $l_m$& \centering $l_n$ - $l_m$ &\centering $l_f$ - $l_n$ &$l_f$\\ [1ex]
				
				\hline
				$\textbf{Case-VII}$:        $l_f$ = $l_n$ > $l_m$
				&\centering $l_m$& \centering $l_n$ - $l_m$ &\centering 0 &$l_f$\\ [1ex]

				\hline
				$\textbf{Case-VIII}$:       $l_f$ = $l_m$ > $l_n$
				&\centering $l_n$& \centering $l_m$ - $l_n$ &\centering 0  &$l_f$\\ [1ex]
				
				\hline
				$\textbf{Case-IX}$:      $l_n$ > $l_f$ = $l_m$
				&\centering $l_m$& \centering 0 &\centering $l_n$ - $l_f$ &$l_n$\\ [1ex]
				
				\hline
				$\textbf{Case-X}$:         $l_n$ = $l_m$ > $l_f$
				&\centering $l_f$& \centering $l_m$ - $l_f$ &\centering 0 &$l_n$\\ [1ex]
				
				\hline
				$\textbf{Case-XI}$:    $l_m$ > $l_f$ = $l_n$
				&\centering $l_n$& \centering 0 &\centering $l_m$ - $l_f$ &$l_m$\\ [1ex]
				
				\hline
				$\textbf{Case-XII}$:     $l_n$ > $l_m$ > $l_f$
				&\centering $l_f$& \centering $l_m$ - $l_f$ &\centering $l_n$ - $l_m$ &$l_n$\\ [1ex]
				
				\hline
				$\textbf{Case-XIII}$:       $l_m$ > $l_f$ > $l_n$
				&\centering $l_n$& \centering $l_f$ - $l_n$ &\centering $l_m$ - $l_f$ &$l_m$\\ [1ex]
				\hline         								
			\end{tabular}
		\end{table*}
		
		The near users decode the data of far and intermediate users to perform SIC. Hence, an index code is designed for near users by considering their respective side information and the index codes of far and intermediate users as additional coded side information.
		The side information at each near user can be modified as \(\mathcal{K}\)$_i^{''}$\ $=$ {\{\(\mathcal{K}\)$_i$ $\cup$ {\(\mathcal{Y}_f \) $\cup$ {\(\mathcal{Y}_m\): $\textit{i}$ $\in$ \(\mathcal{I}\)$_n$\} and the total known set of near users is represented as  \(\mathcal{K}_n^{'} \) $=$ {\{\(\mathcal{K}\)$_i^{''}$: \ $\textit{i}$ $\in$  \(\mathcal{I}\)$_n$\}}. The total want set of near users is represented as  \(\mathcal{W}\)$_n^{'}$\ $=$ {\{\(\mathcal{W}\)$_i$$\setminus${\(\mathcal{W}_i^{c} \): $\textit{i}$ $\in$  \(\mathcal{I}\)$_n$\}}, where \(\mathcal{W}_i^{c} \) is the demand set of near user $V_i$ satisfied through index codes of far users and intermediate users. Hence, the index coding problem for near group users is an ICCSI problem with encoding matrix $\textbf{L}$$_n^{c}$. The index-coded vector is obtained as 
						\begin{equation}
							\label{eqn:near}
							\textbf{y}_n^{c} = \textbf{L}_n^{c}\textbf{x}
						\end{equation}
						where $\textbf{L}$$_n^{c}$ is a $\textit{l}$$_n$$\times$n matrix. The length of the index code designed for near users is denoted as $\textit{l}$$_n$. Let  \(\mathcal{Y}_n\)$=${\{$y$$_n^{c}$, $\textit{i}$ $\in$ [$\textit{l}$$_n$]\}} be the set of index coded packets designed for near users.

						This index code design for near and intermediate users  allows us to reap additional bandwidth savings. Algorithm \ref{alg:1} is developed explicitly for situations where the users can be divided into three groups based on their channel conditions. The channel gain difference between two users of the same group is considered negligible. The channel gain difference between users from different groups is considerably greater.

						We search for maximum, median and minimum channel gains $g_{max}$, $g_{med}$ and $g_{min}$ from a group of $N$ users as shown in line 1 of Algorithm-\ref{alg:1}. Lines 2-14 of Algorithm-\ref{alg:1} show the grouping of users into far, intermediate and near groups depending on the channel conditions. Index code design for far, intermediate and near group users is described in lines 15-18, 19-24 and 25-31 of Algorithm-\ref{alg:1} respectively.
						
						Hence, the conventional IC problem is converted into three IC problems when we group users into three groups. One is a traditional index coding problem for far users, and the others are ICCSI problems for intermediate and near users.
						
\section{Transmissions in 3-Group IC-NOMA system}  
						\label{trans}    
						Depending on the values of $l_{f}$, $l_{m}$ and $l_{n}$, the transmissions in  3-group IC-NOMA scheme can be of the following three types: (i)	\textbf{Conventional IC transmission}, (ii) \textbf{2-Group NOMA transmission}, wherein the index-coded packets designed for two groups are suitably combined at two different power levels, and (iii) \textbf{3-Group NOMA transmission}, wherein the index-coded packets designed for  3 groups are suitably combined at three different power levels.
						For the rest of the discussion, without loss of generality, we assume $\textit{l}$$_f$ > 0, $\textit{l}$$_m$ > 0 and $\textit{l}$$_n$ > 0. If one out of $\textit{l}$$_f$, $\textit{l}$$_m$, $\textit{l}$$_n$ is zero, then the proposed scenario reduces to the IC-NOMA scenario in \cite{PPM}. If both $\textit{l}$$_n$ and $\textit{l}$$_m$ equal zero, the proposed scenario reduces to a Conventional IC scenario for far users.

						Let $S^{IC-NOMA}$ denote the total set of transmitted messages and $l^{IC-NOMA}$ denote the total number of transmissions in the proposed system. Let  $l^{NOMA^{(3)}}$ denote the total number of 3-Group NOMA transmissions, $l^{NOMA^{(2)}}$ denote the total number of 2-Group NOMA transmissions and $l^{IN-IC}$ denotes total number of conventional IC transmissions. Then,

						\begin{equation*}
							S^{IC-NOMA} =  \begin{cases}
								S_{k}^{NOMA^{(3)}} , k \in [l^{NOMA^{(3)}}] \\[1ex]
								
								S_{k}^{NOMA^{(2)}} , k \in [l^{NOMA^{(2)}}]\\[1ex]
								\        								
								S_{k}^{IC}                   ,  k \in [l^{IN-IC}]\\[1ex] 
								
							\end{cases}
						\end{equation*}														
						where, $S_{k}^{NOMA^{(3)}}$ denotes the $k$$^{th}$ 3-Group NOMA transmission, $S_{k}^{NOMA^{(2)}}$ denotes the $k$$^{th}$ 2-Group NOMA transmission and $S_{k}^{IC}$  denotes the $k$$^{th}$ conventional IC transmission.
						
						Let $s_{f_{i}}$$=$ $enc$ ($y_{f_{i}}$), $\forall$ $\textit{i}$ $\in$ [$\textit{l}$$_f]$ ; $s_{m_{i}}$$=$ $enc$ ($y_{m_{i}}^{c}$), $\forall$ $\textit{i}$ $\in$ [$\textit{l}$$_m]$ and  $s_{n_{i}}$$=$ $enc$ ($y_{n_{i}}^{c}$), $\forall$ $\textit{i}$ $\in$ [$\textit{l}$$_n]$. For index codes developed using Algorithm \ref{alg:1}$,$ the transmissions are described by Algorithm \ref{alg:2}.			
									
 The total number of transmissions in the 3-Group IC-NOMA scheme is given as $l^{IC-NOMA}=max(l_{f},l_{m},l_{n})$. Depending on the values of  $l_f$, $l_m$ and $l_n$, the 3-Group IC-NOMA system consists of 13 different cases shown in Table-\ref{Tab:2}

                       \textbf{\textit{Theorem }}1: For a given index coding problem with $\textit{N}$ number of users, the 3-Group IC-NOMA system serves user demands with less than or an equal number of transmissions compared to the conventional IC system.\\
						\textbf{Proof}: For a given index coding problem with $\textit{N}$ number of users, let $l^{IC}$ denote the optimal length of index code in a conventional IC scenario. Considering the same problem as three sub-problems in the 3-Group IC-NOMA scenario, let $l_{f}$ be the length of the index code for far users with $\textit{N}$$_f$ < $\textit{N}$, $l_{m}$ be the length of index code for intermediate users with $\textit{N}$$_m$ < $\textit{N}$  and $l_{n}$ be the length of the index code for near users with $\textit{N}$$_n$ < $\textit{N}$ where $\textit{N}$$_f$ + $\textit{N}$$_m$ + $\textit{N}$$_n$ = $\textit{N}$. Since the conventional index coding problem is split into three sub-problems in a 3-Group IC-NOMA system, we can write $l_{f}$ $\leq$ $l^{IC}$, $l_{m}$ $\leq$ $l^{IC}$ and $l_{n} \leq l^{IC}$ and as mentioned earlier $l^{IC-NOMA}$$=$$max(l_{f},l_{m},l_{n})$.  Hence, $l^{IC-NOMA}$ $\leq$ $l^{IC}$. Thus, the 3-Group IC-NOMA system serves user demands with less than or an equal number of transmissions compared to conventional IC systems. \\
						\\							
                          \textbf{\textit{Theorem }}2: For a given index coding problem with $N$ number of users; $l_{m}$ $\leq$ max($l_{f}$, $l_{n}$) is a sufficient condition for the 3-Group IC-NOMA system to serve user demands with less than or an equal number of transmissions compared to the IC-NOMA system in \cite{PPM}.\\				
						\textbf{Proof}: Consider the IC-NOMA system in \cite{PPM}. let $l_{f}^{*}$ be the length of the  index code designed for  $\textit{N}$$_f^{*}$ far users with $\textit{N}$$_f^{*}$ < $\textit{N}$  and $l_{n}^{*}$ be the length of the index code designed for  $\textit{N}$$_n^{*}$ near users with $\textit{N}$$_n^{*}$ < $\textit{N}$ where $\textit{N}$$_f^{*}$ $+$ $\textit{N}$$_n^{*}$ = $\textit{N}$. Then, the total number of transmissions required by the IC-NOMA system in \cite{PPM} is given by $	l^{*}=max(l_{f}^{*}, l_{n}^{*})$. Considering the same problem as three sub-problems in the 3-Group IC-NOMA scenario, let $l_{f}$ be the length of the index code designed for $\textit{N}$$_f$ far users with $\textit{N}$$_f$ < $\textit{N}$, $l_{m}$ be the length of the index code designed for $\textit{N}$$_m$ intermediate users with $\textit{N}$$_m$ < $\textit{N}$  and $l_{n}$ be the length of the index code designed for  $\textit{N}$$_n$ near users with $\textit{N}$$_n$ < $\textit{N}$ where $\textit{N}$$_f$ + $\textit{N}$$_m$ + $\textit{N}$$_n$ = $\textit{N}$. We have  $\textit{N}$$_f$ $\leq$ $\textit{N}$$_f^{*}$ $\implies$ $\textit{l}$$_f$ $\leq$ $\textit{l}$$_f^{*}$ and $\textit{N}$$_n$ $\leq$ $\textit{N}$$_n^{*}$ $\implies$ $\textit{l}$$_n$ $\leq$ $\textit{l}$$_n^{*}$. So, $max(l_{f}^{*}, l_{n}^{*})$ $\geq$ $max(l_{f}, l_{n})$ $\implies$ $l^{*} \geq max(l_{f}, l_{n})$.
						As mentioned earlier, $l^{IC-NOMA}$$=$$max(l_{f},l_{m},l_{n})$ and for $l_{m}$ $\leq$ max($l_{f}$, $l_{n}$), we have $l^{IC-NOMA}$$=$$max(l_{f},l_{n})$. Therefore, $ l^{IC-NOMA} \leq	l^{*} $.
										
						\begin{example}
							\label{ex:3}
							The server has a set of 7 messages represented as \(\mathcal{X =} \) \{$x_{i}$; $i$ $\in$ [7]\}. Let the initial demand set of users \{$V_{i}$ ; $i$ $\in$ [5]\} before R2V phase be given as  \(\mathcal{D}\) $=$ \{$\{x_{1},x_{4},x_{5},x_{6}\},\{x_{1},x_{5},x_{6}\},\{x_{1},x_{2},x_{6}\}, \{x_{1},x_{3},x_{4}, \\ x_{5},x_{7}\}
							,\{x_{4},x_{7}\}\}$. 
												
							In the R2V phase, the server, upon receiving the demands of each user, sends the total demand set \(\mathcal{X =} \)\{$x_{1},x_{2},x_{3},x_{4},x_{5},x_{6},x_{7}\}$ to the RSU through BS in the direction of vehicle movement. The vehicles entering the RSU range receive data packets as side information, as shown by Table-\ref{Tab:3}. Since the demands of some of the users will be satisfied by side information during the R2V phase, the want set of each user is modified at BS as \(\mathcal{W}\)$_i$= \(\mathcal{D}\)$_i$$\setminus$\{\(\mathcal{D}\)$_i$ $\cap$ \(\mathcal{K}\)$_i$\}.
														         							
							Here, it is assumed that the spacing between $V_1$ and $V_2$ is same as that between $V_2$ and $V_3$ while the spacing between $V_3$ and $V_4$ is longer and same as that between  $V_4$ and $V_5$. The overlap of the side information depends on vehicle spacing; hence, the overlap of side information between $V_3$ and $V_4$ is lesser. Also, $V_5$ enters last, so it could download fewer packets than other vehicles. 
				\vspace{-0.5cm}			
														\begin{center}
																\begin{table}[H]
																		\centering
																		\caption{Known set and modified want set of users (After R2V phase)}
																		\label{Tab:3}
																		\begin{tabular}{|c|c|c|} 
																				\hline
																				User& Known set \(\mathcal{K}\)$_i$& want set \(\mathcal{W}\)$_i$ \\ [0.5ex] 
																				\hline
																				$V_1$ & \{$x_{1},x_{2},x_{3}\}$ & \{$x_{4},x_{5},x_{6}\}$ \\
																				\hline
																				$V_2$ & \{$x_{2},x_{3},x_{4}\}$ & \{$x_{1},x_{5},x_{6}\}$ \\
																				\hline
																				$V_3$ & \{$x_{3},x_{4},x_{5}\}$ & \{$x_{1},x_{2},x_{6}\}$ \\
																				\hline
																				$V_4$ & \{$x_{5},x_{6},x_{7}\}$ & \{$x_{1},x_{3},x_{4}\}$ \\
																				\hline		
																				$V_5$ & \{$x_{7}\}$ & \{$x_{4}\} $\\
																				\hline
																			\end{tabular}
																	\end{table}
															\end{center} 
\vspace{-0.5cm}							
One of the possible set of index coded transmissions  solutions for the above index coding problem are given by $y_1$ = $x_1$ + $x_4$, $y_2$ = $x_2$ + $x_5$, $y_3$ = $x_3$ + $x_6$ and $y_4$ = $x_4$ + $x_7$. The index code length is given by $\textit{l}^{IC}$ = 4. 
							
							Considering the spacing between the vehicles, $V_1$, $V_2$ and $V_3$ are grouped as near users, $V_4$ as intermediate user and $V_5$ as a far user in the 3-Group IC-NOMA system.
							
							For combining designed index codes with NOMA principles, far users must be able to decode desired data without getting the data of near and intermediate users, and intermediate users must be able to decode desired data without getting the data of near users. We cannot use the above index code design in the NOMA scenario as the intermediate user $V_4$ needs the near user's index-coded packet $y_1$ to get the desired message $x_1$. Hence in a 3-Group IC-NOMA scenario, the index code design needs to be modified to reap the benefits of NOMA.
							
							In 3-Group IC-NOMA, to develop index codes that can bring in improved spectral efficiency when combined with the NOMA concept, we consider index coding problem as three different index coding problems, i.e. a conventional IC problem to design index code for far group users and two different ICCSI problems, to design index code for intermediate and near group users respectively.
													
							Based on the known set and want set of far user $V_5$ given in Table-\ref{Tab:3}, 						
							the far user index code is given by $y_{f_{1}}$=$x_4$+$x_7$. The index code length for the far user is $\textit{l}$$_f$$=$1.

							One of the demands of the intermediate user is satisfied by the index code of the far user. Hence the want set and known set of intermediate user $V_4$ given in Table-\ref{Tab:3} is modified as \{$x_5,x_{6},x_{7},x_{4}+x_7\}$ and \{$x_1,x_3\}$.

							Next, an encoding matrix $\textbf{L}$$_m^{c}$ is designed to develop index code for the intermediate user as per (\ref{eqn:inter}) based on the modified known and want sets. Hence, the  intermediate user index code is given by $y_{m_{1}}^{c}$ = $x_1$+$x_7$ and $y_{m_{2}}^{c}$ = $x_3$+$x_6$. The index code length for the intermediate user is $\textit{l}$$_m$$=$2.
							
							The index codes of far and intermediate users satisfy some of the demands of near users. Hence, the want set and known set of near users are modified, as shown in Table-\ref{Tab:5}.
							
							\vspace{-0.6cm}
							\begin{center}
								\begin{table}[H]
									\small
									\caption{IC NOMA problem for near users}
									\label{Tab:5}
									\begin{tabular}{|p{1.2cm} |p{5.2cm}|p{1.25cm}|}
										\hline
										User& Known set& Want set\\ [0.5ex] 
										\hline
										$V_1$ &\{$x_1,x_{2},x_{3},x_{4}+x_7,x_{1}+x_7,x_{3}+x_6\}$ & \{$x_5\}$ \\[0.2ex] 
										\hline
										$V_2$ &\{$x_2,x_{3},x_{4},x_{4}+x_7,x_{1}+x_7,x_{3}+x_6\}$ & \{$x_5\}$ \\[0.2ex] 
										\hline
										$V_3$ &\{$x_3,x_{4},x_{5},x_{4}+x_7,x_{1}+x_7,x_{3}+x_6\}$ & \{$x_2\}$ \\[0.2ex] 
										\hline             		
									\end{tabular}
								\end{table}
							\end{center} 
							\vspace{-0.3cm}
							An encoding matrix $\textbf{L}$$_n^{c}$ is designed to develop index code for near users as per (\ref{eqn:near}) based on Table-\ref{Tab:5}. Hence, the index code designed for near users is given by $y_{n_{1}}^{c}$ = $x_2$+$x_5$. The index code length for near users is $\textit{l}$$_n$$=$1.

							Let $s_{n_{1}}$$=$ enc($y_{n_{1}}^{c}$), $s_{m_{1}}$$=$enc($y_{m_{1}}^{c}$) , $s_{m_{2}}$$=$enc($y_{m_{2}}^{c}$) and $s_{f_{1}}$$=$enc($y_{f_{1}}$). The designed index codes are combined using NOMA principles in the 3-Group IC-NOMA system using Algorithm-\ref{alg:2}. Hence, the transmitted signals are $S_{1}^{NOMA^{(3)}}$$=$$\sqrt{\alpha P}$$s_{n_{1}}$$+\sqrt{\beta P}$$s_{m_{1}}$$+\sqrt{\gamma P}$$s_{f_{1}}$, $S_{1}^{IC}$$=$$\sqrt{P}$$s_{m_{2}}$, where 
							$\alpha$<$\beta$<$\gamma$ and $\alpha$+$\beta$+$\gamma$$=$1.

							The far users receiving $S_{1}^{NOMA^{(3)}}$  decode $s_{f_{1}}$ considering interference due to $s_{m_{1}}$ and $s_{n_{1}}$ as noise. The intermediate users receiving  $S_{1}^{NOMA^{(3)}}$ gets the far user data 
							$s_{f_{1}}$  and performs SIC to get $s_{m_{1}}$  and from $S_{1}^{IC}$ they decode $s_{m_{2}}$. The near users receiving $S_{1}^{NOMA^{(3)}}$ decode $s_{f_{1}}$ and $s_{m_{1}}$, then they perform SIC to get $s_{n_{1}}$. Hence, far users get  \{$s_{f_{1}}$\}, intermediate users get \{$s_{m_{1}}$,$s_{m_{2}}$\} and near users get \{$s_{n_{1}}$\}. Hence, in Example \ref{ex:3}, the total number of transmissions in the 3-Group IC-NOMA scheme is $l^{IC-NOMA}$=2, out of which one is a 3-Group NOMA transmission, and the other is a conventional IC transmission.
						\end{example}
						
							\begin{example}
							\label{ex:4}
							The server has a set of 7 messages represented as \(\mathcal{X =} \) \{$x_{i}$; $i$ $\in$ [7]\}. Let the initial demand set of users \{$V_{i}$ ; $i$ $\in$ [6]\} before R2V phase be given as  \(\mathcal{D}\) = \{$\{x_{1},x_{2},x_{4},x_{5},x_{6}\},\{x_{1},x_{2},x_{5},x_{6}\},\{x_{1},x_{2},x_{3}\},\\ \{x_{3},x_{4},  x_{6},x_{7}\}
							,\{x_{4},x_{7}\}, \{x_{1},x_{7}\}\}$. \\
							\vspace{-0.3cm}
							
							In the R2V phase, the server, upon receiving the demands of each user, sends the total demand set \(\mathcal{X =} \)\{$x_{1},x_{2},x_{3},x_{4},x_{5},x_{6},x_{7}\}$ to the RSU through BS in the direction of vehicle movement. The vehicles entering the RSU range receive data packets as side information, as shown by Table - \ref{Tab:6}. Since the demands of some of the users will be satisfied by side information during the R2V phase, the want set of each user is modified at BS as \(\mathcal{W}\)$_i$= \(\mathcal{D}\)$_i$$\setminus$\{\(\mathcal{D}\)$_i$ $\cap$ \(\mathcal{K}\)$_i$\}.\\
							\vspace{-0.5cm}						
							\begin{center}
								\begin{table}[H]         								
									\centering
									\caption{Known set and modified want set of users (After R2V phase)}
									\label{Tab:6}
									\begin{tabular}{|c|c|c|} 
										\hline
										User& Known set  \(\mathcal{K}\)$_i$& want set \(\mathcal{W}\)$_i$ \\ [0.5ex] 
										\hline
										$V_1$ & \{$x_{1},x_{2},x_{3}\}$ & \{$x_{4},x_{5},x_{6}\}$ \\
										\hline
										$V_2$ & \{$x_{2},x_{3},x_{4}\}$ & \{$x_{1},x_{5},x_{6}\}$ \\
										\hline
										$V_3$ & \{$x_{4},x_{5},x_{6}\}$ & \{$x_{1},x_{2},x_{3}\}$ \\
										\hline
										$V_4$ & \{$x_{5},x_{6},x_{7}\}$ & \{$x_{3},x_{4}\}$ \\
										\hline		
										$V_5$ & \{$x_{7}\}$ & \{$x_{4}\} $\\
										\hline
										$V_6$ & \{$x_{7}\}$ & \{$x_{1}\} $\\
										\hline
									\end{tabular}
								\end{table}
							\end{center} 
							\vspace{-0.4cm}
							Here, it is assumed that the spacing between $V_1$ and $V_2$ is the same as between $V_3$ and $V_4$. $V_5$ and $V_6$ enter last, so they could download fewer packets than other vehicles.
							
							The encoding matrix $\textbf{L}$ representing one of the possible linear solutions for the above index coding problem is \\         							
							$\textbf{L} =\begin{bmatrix}
								1 & 0 & 0& 1 &0&0&0\\
								0& 1 & 0& 0 &1&0&0\\
								0 & 0 & 1& 0 &0&1&0\\
								0& 0 & 0& 1 &0&0&1\\
							\end{bmatrix}$ and the corresponding index-coded transmissions are given by $y_1$ = $x_1$ + $x_4$, $y_2$ = $x_2$ + $x_5$, $y_3$ = $x_3$ + $x_6$ and $y_4$ = $x_4$ + $x_7$. The index code length is given by $\textit{l}^{IC}$ = 4. 
							The decoding process at different users is given in Table-\ref{Tab:7}.\\
							\vspace{-0.7cm}
							\begin{center}
								\begin{table}[H]         								
									\centering
									\caption{Decoding process at receivers}
									\label{Tab:7}
									\begin{tabular}{|c|c|c|}
										\hline
										Receiver& Known set& Decoding process\\ [0.5ex] 
										\hline
										$V_1$ &\{$x_{1},x_{2},x_{3}\}$& $y_1$+$x_1$=$x_4$,$y_2$+$x_2$=$x_5$,$y_3$+$x_3$=$x_6$\\[0.2ex] 
										\hline
										$V_2$ &\{$x_{2},x_{3},x_{4}\}$& $y_1$+$x_4$=$x_1$,$y_2$+$x_2$=$x_5$,$y_3$+$x_3$=$x_6$\\[0.2ex] 
										\hline
										$V_3$ &\{$x_{3},x_{4},x_{5}\}$& $y_1$+$x_4$=$x_1$,$y_2$+$x_5$=$x_2$,$y_3$+$x_6$=$x_3$\\[0.2ex] 
										\hline
										$V_4$ &\{$x_{5},x_{6},x_{7}\}$& $y_3$+$x_6$=$x_3$,$y_4$+$x_7$=$x_4$\\[0.2ex] 
										\hline		
										$V_5$ &\{$x_{7}\}$& $y_4$+$x_7$=$x_4$\\[0.2ex] 
										\hline
										$V_6$ &\{$x_{7}\}$& $y_1$+$y_4$+$x_7$=$x_1$\\[0.2ex] 
										\hline      		
									\end{tabular}
								\end{table}
							\end{center} 
							\vspace{-0.7cm}
							
							We group $V_1$ and $V_2$ as near users, $V_3$ and $V_4$ as intermediate users and $V_5$ and $V_6$ as far users. We cannot use the above index code design in the NOMA scenario as the intermediate user $V_3$ needs the near user's index-coded packet $y_1$ to get desired message $x_1$. Hence, in a 3-Group IC-NOMA scenario, the index code design needs to be modified for combining with NOMA.\\.
							
							Based on the known set and want set of far users $V_5$ and $V_6$ given in Table-\ref{Tab:6}, an encoding matrix $\textbf{L}$$_f$ is designed to develop index code for far users as per (\ref{eqn:far}).  Hence, the encoding matrix $\textbf{L}$$_f$ corresponding to far users is designed as  $\textbf{L}_f =\begin{bmatrix}
								1& 0 & 0& 0&0&0&1\\
								0& 0 & 0& 1 &0&0&1\\        							         								
							\end{bmatrix}.$    							The far user index code is given by $y_{f_{1}}$$=$$x_1$+$x_7$ and $y_{f_{2}}$$=$$x_4$+$x_7$. The index code length for the far user is $\textit{l}$$_f$$=$2.

									

							Some of the demands of intermediate users are satisfied by the index code of the far user. Hence the want set and known set of intermediate users $V_3$ and $V_4$ given in Table-\ref{Tab:6} is modified as \{$\{x_{4},x_{5},x_{6},x_{1+}x_7,x_{4+}x_7\},\{x_{5},x_{6},x_{7},x_{1+}x_{7},x_{4+}x_7\}\}$ and \{$\{x_{2},x_{3}\},x_{3}\}$
							
							%
							Next, an encoding matrix $\textbf{L}$$_m^{c}$ is designed to develop index code for intermediate users as per (\ref{eqn:inter}) based on the modified known and want sets of the intermediate users. Hence, the encoding matrix $\textbf{L}$$_m^{c}$ corresponding to intermediate users is designed as	$\textbf{L}_m^{c} =\begin{bmatrix}
								0& 1 & 0& 0 &1&0&0\\
								0& 0 & 1& 0 &0&1&0\\
							\end{bmatrix}.$        							The intermediate users index code is given by $y_{m_{1}}^{c}$=$x_2$+$x_5$ and $y_{m_{2}}^{c}$=$x_3$+$x_6$. The index code length for intermediate users is $\textit{l}$$_m$$=$2.
							
							All near-users' demands are satisfied when they exploit their side information and the index codes of far and intermediate users. So, $\textit{l}$$_n$$=$0.

							Let  $s_{m_{1}}$$=$enc($y_{m_{1}}^{c}$), $s_{m_{2}}$$=$enc($y_{m_{2}}^{c}$), $s_{f_{1}}$=enc($y_{f_{1}}$) and $s_{f_{2}}$=enc($y_{f_{2}}$). The designed index codes are combined using NOMA principles in the 3-Group IC-NOMA system using Algorithm-\ref{alg:2}.

							Hence, the transmitted signals are $S_{1}^{NOMA^{(2)}}$$=$$\sqrt{\alpha_{1} P}$$s_{m_{1}}$+$\sqrt{(1-\alpha_{1})P}$$s_{f_{1}}$, $S_{2}^{NOMA^{(2)}}$$=$$\sqrt{\alpha_{1} P}$$s_{m_{2}}$+$\sqrt{(1-\alpha_{1})P}$$s_{f_{2}}$, \\ where $\alpha_{1}$ < 0.5.

							The far users receiving $S_{1}^{NOMA^{(2)}}$  decode $s_{f_{1}}$ considering interference due to $s_{m_{1}}$ as noise and on receiving $S_{2}^{NOMA^{(2)}}$  decode $s_{f_{2}}$ considering interference due to $s_{m_{2}}$ as noise. The intermediate users receving $S_{1}^{NOMA^{(2)}}$ get the far user data 
							$s_{f_{1}}$  and perform SIC to get $s_{m_{1}}$ and on receving $S_{2}^{NOMA^{(2)}}$ get the far user data 
							$s_{f_{2}}$  and perform SIC to get $s_{m_{2}}$. The near users receiving $S_{1}^{NOMA^{(2)}}$ decode $s_{f_{1}}$ and $s_{m_{1}}$ and on receiving $S_{2}^{NOMA^{(2)}}$ decode $s_{f_{2}}$ and $s_{m_{2}}$. Hence far users get  \{$s_{f_{1}},s_{f_{2}}$\}, intermediate users get \{$s_{m_{1}},s_{m_{2}}$\},  and near users get \{$s_{f_{1}},s_{f_{2}},s_{m_{1}},s_{m_{2}}$\}.	Hence, in Example \ref{ex:4}, the total number of transmissions in the 3-Group IC-NOMA scheme is $l^{IC-NOMA}$=2 where both the transmissions are 2-Group NOMA transmissions.
							
						\end{example}

\section{Rate Analysis}
\label{RA}
In this section, we quantify the sum rate of the transmission scheme and improvement in the achievable information rate of the  3-Group IC-NOMA transmission scheme over the conventional IC system. Let $g_n$$ = $min($g_i$); \textit{i} $\in$ \(\mathcal{I}\)$_n$, $g_m$$ = $min($g_i$); \textit{i} $\in$ \(\mathcal{I}\)$_m$ and $g_f$$ = $min($g_i$); \textit{i} $\in$ \(\mathcal{I}\)$_f$ denote the channel gain of users with worst channel conditions in near, intermediate and far groups respectively. $g_n$, $g_m$, and $g_f$ determine the minimum achievable rate for near, intermediate and far group users. 

We perform rate analysis at each user, assuming additive white Gaussian noise of zero mean and unit variance. All rates in further discussion are expressed in bits per channel use (bpcu). In all further discussions, the term scheme is used instead of the 3-Group IC-NOMA transmission scheme.

\subsection{\textit{Convetional IC Scenario}}
In the conventional IC system, let the power required for each transmission be denoted as $P$ and the number of index-coded transmissions required to satisfy the demands of all users be denoted as $l^{IC}$.
Thus, the transmitted messages  are $S_{i}^{IC} = \sqrt{P}enc(y^{i}) , i=[l^{IC}]$ ;
where $y^{i}$ denotes the $i^{th}$ index-coded packet.
The minimum achievable rate that all users in the near, intermediate and far groups could achieve is denoted as  $R_{n}^{IC}$, $R_{m}^{IC}$ and $R_{f}^{IC}$ respectively. Then,
\begin{equation}
	\label{1}
	R_{n}^{IC} = log_2(1+g_nP)
\end{equation}
\begin{equation}
	\label{2}
	R_{m}^{IC} = log_2(1+g_mP)
\end{equation}
\begin{equation}
	\label{3}
	R_{f}^{IC} = log_2(1+g_fP).
\end{equation}
Since $g_n > g_m > g_f$, we have $R_{n}^{IC} > R_{m}^{IC} > R_{f}^{IC}$.  Let $R^{IC}$ denote the minimum achievable rate by all users. Then,
\begin{equation}
	\label{4}
	R^{IC} = log_2(1+g_fP)
\end{equation}      							       							
\subsection{\textit{IC-NOMA Scenario}}
This section considers Case-I, Case-II, Case-III and  Case-IV of the scheme given in Table-\ref{Tab:2}. Rate analysis for all other cases is provided in Appendix-1. For comparison, the power per transmission is considered the same as that of a conventional index-coded system, i.e.,$P$.

1) $\textit{CASE}$ I ($l_f=l_m=l_n$) : In this case, a total of $l_f$ 3-Group NOMA transmissions will be required by the  scheme. The transmitted messages are $	S_{k}^{IC-NOMA} = 
\sqrt{\alpha P}s_{n_{k}} +\sqrt{\beta P}s_{m_{k}}+\sqrt{\gamma P}s_{f_{k}}, k \in [ l_f ]$         							
with $\alpha<\beta<\gamma$ and $\alpha+\beta+\gamma=1$.

Let $R_{f}^{IN-NOMA^{(3)}}$ denote the minimum achievable rate that could be achieved by users in the far group from a 3-Group NOMA transmission. Then, $	R_{f}^{IN-NOMA^{(3)}} = log_2 \left (1+ \frac{\gamma P g_f}{1 + (\alpha + \beta) P g_f} \right)$.\vspace{0.1cm}

Intermediate users perform SIC for far users' data. Let $R_{m}^{IN-NOMA^{(3)}}$ denote the minimum achievable rate that could be achieved by users in the intermediate group from a 3-Group NOMA transmission. Then, $R_{m}^{IN-NOMA^{(3)}} = log_2 \left (1+ \frac{\beta P g_m}{1 + \alpha P g_m} \right)$. \vspace{0.1cm}

Near users perform SIC for far and intermediate users' data. Let $R_{n}^{IN-NOMA^{(3)}}$ denote the minimum achievable rate that could be achieved by users in the near group from a  3-Group NOMA transmission. Then, $	R_{n}^{IN-NOMA^{(3)}} = log_2 \left (1+ {\alpha P g_n} \right)$.

The sum rate of transmission scheme is denoted as $R^{IN-NOMA^{(3)}}$ for a single 3-Group NOMA transmission and is given by

\begin{align}        
	\fontsize{8}{8}			
	\begin{split}	
		\label{5}				
		R^{IN-NOMA^{(3)}} 
		=& log_2 \left (\frac{(1 + P g_f)(1 + (\alpha + \beta) P g_m)(1+ {\alpha P g_n})}{(1 + (\alpha + \beta) P g_f)(1+ {\alpha P g_m})}\right)       								
	\end{split}
\end{align}


Consider (\ref{4}) and (\ref{5}) to analyse the improvement in rate the scheme can support from a single 3-Group NOMA transmission.
\begin{equation}
	\label{6}
	\fontsize{8}{8}			
	\begin{split}
		R^{IN-NOMA^{(3)}} - R^{IC}=log_2 \left\{\frac{(1+(\alpha+\beta)Pg_m)(1+{\alpha Pg_n})}{(1+(\alpha+\beta) Pg_f)(1+ {\alpha Pg_m})}\right\}
	\end{split}
\end{equation}
Since $g_n> g_m> g_f$, we have   $R^{IN-NOMA^{(3)}}$ > $ R^{IC}$. Hence, the 3-Group NOMA transmission can support a higher information rate than a conventional IC system.

2) $\textit{CASE}$ II ($l_f>l_m>l_n$) : In this case, a total of $l_f$  transmissions will be required by the  scheme,  with $l_n$  3-Group NOMA transmissions, $l_m-l_n$ 2-Group NOMA transmissions and $l_f-l_m$  index-coded transmissions. The transmitted messages are
\vspace{-0.2cm}
\begin{equation*}
	\fontsize{8}{8}
	S_{k}^{IC-NOMA} = \begin{cases}
		
		\sqrt{\alpha P}s_{n_{k}} +\sqrt{\beta P}s_{m_{k}}+\sqrt{\gamma P}s_{f_{k}}, k \in [l_n] \\[0.5ex]
		
		\sqrt{\alpha_{1} P}s_{m_{k}}+\sqrt{(1-\alpha_{1})P}s_{f_{k}},k=\{l_n+1,\cdots l_m\}\\[0.5ex]
		and\\
		\sqrt{P}s_{f_{k}}                 ,  k=\{l_m+1,...l_f\}\\[0.5ex]
	\end{cases}
\end{equation*}         							         							
with $\alpha<\beta<\gamma$ , $\alpha+\beta+\gamma=$1  and $\alpha_{1}$ < 0.5. The sum rate of the transmission scheme for a 3-Group NOMA transmission will be the same as (\ref{5}).

In this case, let $R_{f,(m,f)}^{IN-NOMA^{(2)}}$  and $R_{m,(m,f)}^{IN-NOMA^{(2)}}$ \vspace{0.1cm}  respectively denote the minimum information rate that users could achieve in far and intermediate groups from a 2-Group NOMA transmission. Then,
\vspace{-0.2cm}
\begin{equation}    
	\label{fmf}     								
	R_{f,(m,f)}^{IN-NOMA^{(2)}}=log_2 \left (1+ \frac{(1- \alpha_{1}) P g_f}{1 + \alpha_{1}  P g_f} \right)          								
\end{equation}
\vspace{-0.2cm}
\begin{equation}
	\label{mmf}        								
	R_{m,(m,f)}^{IN-NOMA^{(2)}}=log_2 \left (1+ {\alpha_{1} P g_m} \right).           								
\end{equation}

The sum rate of the transmission scheme for a  2-Group NOMA transmission where the index-coded packets of far and intermediate group users are combined is denoted as $R_{(m,f)}^{IN-NOMA^{(2)}}$ and is given by \vspace{0.1cm}
\begin{equation}
	\label{7}
	\begin{split}
		R_{(m,f)}^{IN-NOMA^{(2)}}  					
		& = log_2 \left (\frac{(1 + P g_f)(1 + \alpha_{1} P g_m)}{(1 + \alpha_{1}   P g_f)} \right)
	\end{split}
\end{equation}

Consider (\ref{4}) and (\ref{7}) to analyse the improvement in the rate that can be supported by a 2-Group NOMA transmission in this case.
\vspace{-0.2cm}
\begin{equation}
	\label{8}
	\begin{split}
		R_{(m,f)}^{IN-NOMA^{(2)}} - R^{IC} = log_2 \left\{ \frac{1 + \alpha_{1}  P g_m}{1+ {\alpha_{1} P g_f}} \right\}
	\end{split}
\end{equation}

Since $g_m>g_f$, we have $R_{(m,f)}^{IN-NOMA^{(2)}}$ > $R^{IC}$. Hence, the 2-Group NOMA transmission where index coded packets designed for intermediate and far users are combined  can support a higher information rate than a conventional IC system. In this case, the index-coded transmissions are index-coded packets designed for far users. Hence, the achievable rate for IC transmissions is given by (\ref{3}).

\begin{table*}[htbp]	
	\centering
	\begin{tabular}{p{15cm}}
		\begin{equation}
			\begin{split}
				\label{18}
				\zeta =\left (\frac{(1 +P_ag_f)(\alpha P_a(g_n-g_m)+(\alpha+\beta)P_a(g_m-g_f)+\alpha(\alpha+\beta)P_{a}^{2}g_m(g_n-g_f))}{g_f(1+\alpha P_ag_m+(\alpha+\beta)P_ag_f+\alpha(\alpha+\beta)P_{a}^{2}g_mg_f)}\right)
			\end{split}
		\end{equation}				
	\end{tabular}
\end{table*}

3) $\textit{CASE}$ III ($l_n>l_f>l_m$) : In this case, a total of $l_n$ transmissions will be required by the scheme, with $l_m$  3-Group NOMA transmissions, $l_f-l_m$ 2-Group NOMA transmissions and $l_n-l_f$ index-coded transmissions. The transmitted messages are
\vspace{-0.2cm}
\begin{equation*}
	\fontsize{8}{8}
	S_{k}^{IC-NOMA} = \begin{cases}
		
		\sqrt{\alpha P}s_{n_{k}} +\sqrt{\beta P}s_{m_{k}}+\sqrt{\gamma P}s_{f_{k}}, k \in [l_m] \\[1ex]
		
		\sqrt{\alpha_{1} P}s_{n_{k}}+\sqrt{(1-\alpha_{1})P}s_{f_{k}},k=\{l_m+1,\cdots l_f\}\\[1ex]
		and\\
		\sqrt{P}s_{n_{k}}                 ,  k=\{l_f+1,\cdots l_n\}\\[1ex]
	\end{cases}
\end{equation*}
with $\alpha<\beta<\gamma$, $\alpha+\beta+\gamma=$1  and $\alpha_{1}$ < 0.5. The sum rate of the transmission scheme for a 3-Group NOMA transmission will be the same as (\ref{5}).\vspace{0.05cm}

In this case, let $R_{f,(n,f)}^{IN-NOMA^{(2)}}$  and $R_{n,(n,f)}^{IN-NOMA^{(2)}}$ \vspace{0.1cm} respectively denote the minimum information rate that users could achieve in far and near groups from a 2-Group NOMA transmission. Then,
\vspace{-0.2cm}
\begin{equation}  
	\label{fnf}       								
	R_{f,(n,f)}^{IN-NOMA^{(2)}} = log_2 \left (1+ \frac{(1- \alpha_{1}) P g_f}{1 + \alpha_{1}  P g_f} \right)           								
\end{equation}
\vspace{-0.2cm}
\begin{equation}        
	\label{nnf} 								
	R_{n,(n,f)}^{IN-NOMA^{(2)}}=log_2 \left (1+ {\alpha_{1} P g_n} \right).          								
\end{equation}

The sum rate of the transmission scheme for a  2-Group NOMA transmission where the index-coded packets of far and near group users are combined is denoted as $R_{(n,f)}^{IN-NOMA^{(2)}}$  and is given by \vspace{0.1cm}
\vspace{-0.2cm}
\begin{equation}
	\begin{split}
		\label{9}
		R_{(n,f)}^{IN-NOMA^{(2)}}
		&= log_2 \left (\frac{(1 + P g_f)(1 + \alpha_{1} P g_n)}{(1 + \alpha_{1}   P g_f)} \right).
	\end{split}
\end{equation}

Consider (\ref{4}) and (\ref{9}) to analyse the improvement in the rate that can be supported by a  2-Group NOMA transmission in this case.
\vspace{-0.2cm}
\begin{equation}
	\begin{split}
		\label{10}
		R_{(n,f)}^{IN-NOMA^{(2)}} - R^{IC} = log_2 \left\{ \frac{1 + \alpha_{1}  P g_n}{1+ {\alpha_{1} P g_f}} \right\}
	\end{split}
\end{equation}
Since $g_n$ > $g_f$, we have $R_{(n,f)}^{IN-NOMA^{(2)}}$ > $R^{IC}$.  Hence, the 2-Group NOMA transmission where index coded packets designed for near and far users are combined  can support a higher information rate than a conventional IC system.

In this case, the index-coded transmissions are index-coded packets designed for near users. Hence, 
the achievable rate for IC transmissions is given by (\ref{1}).

Consider (\ref{1}) and (\ref{4}) to analyse the improvement in the rate that the conventional IC transmissions, in this case, can support.         
\vspace{-0.2cm}							
\begin{equation}
	\label{11}
	\begin{split}
		R_n^{IN-IC} - R^{IC} = log_2 \left\{ \frac{1 + P g_n}{1+ {P g_f}} \right\}.
	\end{split}
\end{equation}
Since $g_n$ > $g_f$, we have $R_n^{IN-IC}$ > $R^{IC}$. Hence, IC transmissions meant for near users   can support a higher information rate than a conventional IC system.

4) $\textit{CASE}$ IV ($l_m > l_n> l_f$) : In this case, a total of $l_m$ transmissions will be required by the scheme, with $l_f$ 3-Group NOMA transmissions, $l_n-l_f$  2-Group NOMA transmissions and $l_m-l_n$ number of index-coded transmissions. The transmitted messages are
\vspace{-0.2cm}
\begin{equation*}
	\fontsize{8}{8}
	S_{k}^{IC-NOMA} = \begin{cases}
		
		\sqrt{\alpha P}s_{n_{k}} +\sqrt{\beta P}s_{m_{k}}+\sqrt{\gamma P}s_{f_{k}}, k \in [l_f] \\[0.5ex]
		
		\sqrt{\alpha_{1} P}s_{n_{k}}+\sqrt{(1-\alpha_{1})P}s_{m_{k}},k=\{l_f+1,\cdots l_n\}\\[0.5ex]
		and\\
		\sqrt{P}s_{m_{k}}                 ,  k=\{l_n+1,\cdots l_m\}\\[0.5ex]
	\end{cases}
\end{equation*}
with $\alpha<\beta<\gamma$, $\alpha+\beta+\gamma$=1  and $\alpha_{1}$ < 0.5. The sum rate of the transmission scheme for a 3-Group NOMA transmission is the same as (\ref{5}).\vspace{0.05cm}

In this case, let $R_{m,(n,m)}^{IN-NOMA^{(2)}}$  and $R_{n,(n,m)}^{IN-NOMA^{(2)}}$ \vspace{0.1cm} respectively denote the minimum information rate that could be achieved by users in intermediate and near groups from  a 2-Group NOMA transmission. Then,
\vspace{-0.2cm}
\begin{equation}        
	\label{mnm}								
	R_{m,(n,m)}^{IN-NOMA^{(2)}} = log_2 \left (1+ \frac{(1- \alpha_{1}) P g_m}{1 + \alpha_{1}  P g_m} \right)          								
\end{equation}
\vspace{-0.2cm}
\begin{equation}     
	\label{nnm}    								
	R_{n,(n,m)}^{IN-NOMA^{(2)}}=log_2 \left (1+ {\alpha_{1} P g_n} \right).          								
\end{equation}

The sum rate of the transmission scheme of a single 2-Group NOMA transmission where the IC packets of near and intermediate group users are combined is denoted as $R_{(n,m)}^{IN-NOMA^{(2)}}$  and is given by \vspace{0.1cm}
\vspace{-0.2cm}
\begin{equation}
	\label{13}
	\begin{split}
		R_{(n,m)}^{IN-NOMA^{(2)}} 
		&= log_2 \left (\frac{(1 + P g_m)(1 + \alpha_{1} P g_n)}{(1 + \alpha_{1}   P g_m)} \right).
	\end{split}
\end{equation}

Consider (\ref{4}) and  (\ref{13}) to analyse the improvement in the rate that can be supported by a 2-Group NOMA transmission in this case.
\vspace{-0.2cm}
\begin{equation}
	\begin{split}
		\label{14}
		R_{(n,m)}^{IN-NOMA^{(2)}} - R^{IC} = log_2 \left\{ \frac{(1 + P g_m)(1 + \alpha_{1}  P g_n)}{(1 + P g_f)(1 + \alpha_{1}  P g_m)} \right\}
	\end{split}
\end{equation}
Since $g_n$ > $g_m$ > $g_f$, we have $R_{(n,m)}^{IN-NOMA^{(2)}}$ > $R^{IC}$.      Hence, the 2-Group NOMA transmission where index coded packets designed for near and intermediate users are combined  can support a higher information rate than a conventional IC system.  					
		
In this case, the index-coded transmissions are index-coded packets designed for intermediate users. Hence, 
the achievable rate for IC transmissions is given by (\ref{2}).

Consider (\ref{2}) and (\ref{4}) to analyse the improvement in the rate that conventional IC transmissions, in this case, can support.
\vspace{-0.2cm}
\begin{equation}
	\begin{split}
		\label{15}
		R_m^{IN-IC} - R^{IC} = log_2 \left\{ \frac{1 + P g_m}{1+ {P g_f}} \right\}
	\end{split}
\end{equation}
Since $g_m$ > $g_f$, we have 	$R_m^{IN-IC}$ > $R^{IC}$. Hence, IC transmissions meant for intermediate users   can support a higher information rate than a conventional IC system.

It is important to note that the rate analysis of these four cases is sufficient to conclude that except for conventional IC transmission meant for far users, all the transmissions in the scheme can support a higher sum rate of the transmission scheme when compared to a traditional IC system.

\section{Transmission Power Analysis}

\begin{table*}[htbp]
	\centering   				
	\caption{Power savings in each case of 3-Group IC-NOMA scheme}
	\label{Tab:8}
	\begin{tabular}{|p{4.05cm} |p{8.4cm}|}
		\hline
		
		$\textbf{Case}$   
		& $P_{saving}^{(i)}$\\ [1.5ex] 
		
		\hline
		
		$\textbf{Case-I}$:        $l_f$ = $l_m$ = $l_n$
		&$(l^{IC}$$- l_f)$$P^{IC}$$ + \zeta l_f$\\ [1.5ex]
		
		\hline
		$\textbf{Case-II}$:       $l_f$ > $l_m$ > $l_n$
		&$(l^{IC}-l_{m})P^{IC} + \zeta l_n  + (l_m-l_n)\zeta_{(m,f)}$\\[1.5ex]
		
		\hline
		$\textbf{Case-III}$:        $l_n$ > $l_f$ > $l_m$
		&$(l^{IC}-l_{n})P^{IC} + \zeta l_m  + (l_f-l_m)\zeta_{(n,f)} + (l_n-l_f)\zeta_{(n)}$\\[1.5ex]
		
		\hline
		$\textbf{Case-IV}$:     $l_m$ > $l_n$ > $l_f$
		&$(l^{IC}-l_{m})P^{IC} + \zeta l_f  + (l_n-l_f)\zeta_{(n,m)} + (l_m-l_n)\zeta_{(m)}$\\[1.5ex]
		\hline

		$\textbf{Case-V}$:        $l_f$ > $l_m$ = $l_n$
		&$(l^{IC}-l_{n})P^{IC} + \zeta l_n  $\\[1.5ex]
		
		\hline
		$\textbf{Case-VI}$:        $l_f$ > $l_n$ > $l_m$
		&$(l^{IC}-l_{n})P^{IC} + \zeta l_m  + (l_n-l_m)\zeta_{(n,f)} $\\[1.5ex]
		
		\hline
		$\textbf{Case-VII}$:        $l_f$ = $l_n$ > $l_m$
		&$(l^{IC}-l_{n})P^{IC} + \zeta l_m  + (l_n-l_m)\zeta_{(n,f)}$\\[1.5ex]

		\hline
		$\textbf{Case-VIII}$:       $l_f$ = $l_m$ > $l_n$
		&$(l^{IC}-l_{m})P^{IC} + \zeta l_n  + (l_m-l_n)\zeta_{(m,f)} $\\[1.5ex]
		
		\hline
		$\textbf{Case-IX}$:      $l_n$ > $l_f$ = $l_m$
		&$(l^{IC}-l_{n})P^{IC} + \zeta l_m  +  (l_n-l_f)\zeta_{(n)}$\\[1.5ex]
		
		\hline
		$\textbf{Case-X}$:         $l_n$ = $l_m$ > $l_f$
		&$(l^{IC}-l_{m})P^{IC} + \zeta l_f  + (l_m-l_f)\zeta_{(n,m)} $\\[1.5ex]
		
		\hline
		$\textbf{Case-XI}$:    $l_m$ > $l_f$ = $l_n$
		&$(l^{IC}-l_{m})P^{IC} + \zeta l_n  + (l_m-l_f)\zeta_{(m)}$\\[1.5ex]
		
		\hline
		$\textbf{Case-XII}$:     $l_n$ > $l_m$ > $l_f$
		&$(l^{IC}-l_{n})P^{IC} + \zeta l_f  + (l_m-l_f)\zeta_{(n,m)} + (l_n-l_m)\zeta_{(n)}$\\[1.5ex]
		
		\hline
		$\textbf{Case-XIII}$:       $l_m$ > $l_f$ > $l_n$
		&$(l^{IC}-l_{m})P^{IC} + \zeta l_n  + (l_f-l_n)\zeta_{(m,f)} + (l_m-l_f)\zeta_{(m)}$\\[1.5ex]
		\hline         								
	\end{tabular}
\end{table*}

In this section, we demonstrate how the scheme outperforms traditional IC  in terms of power consumption. We evaluate the power per transmission required for the scheme to provide an achievable information rate at least as good as that of conventional IC. We also compute the total power consumption for conventional IC and 3-Group IC-NOMA schemes to quantify the power savings of the scheme. This section considers Case-I, Case-II, Case-III and  Case-IV of the scheme. Power analysis for all other cases is given in Appendix-2.

Considering  $P^{IC}$ as the power per transmission for conventional IC, (\ref{4}) is modified as 
\vspace{-0.2cm}
\begin{equation}
	\label{16}
	R^{IC} = log_2(1 + g_f P^{IC})
\end{equation}

1) $\textit{CASE}$ I  ($l_f=l_m=l_n$) : Considering $P_{a}$ as power per 3-Group NOMA transmission, (\ref{5}) is modified as

\begin{align}
	\fontsize{8}{8}
	\begin{split}
		\label{17}
		R^{IN-NOMA^{(3)}}=
		log_2 \left(\frac{(1 + P_{a} g_f)(1 + (\alpha + \beta) P_{a} g_m)(1+ {\alpha P_{a} g_n})}{(1 + (\alpha + \beta) P_{a} g_f)(1+ {\alpha P_{a} g_m})}\right).
	\end{split} 
\end{align}

To find the minimum power requirement for 3-Group IC-NOMA transmission that makes  $R^{IN-NOMA^{(3)}}$ at least as good as $R^{IC}$, consider (\ref{16}) and (\ref{17}).
\vspace{-0.2cm}
\begin{align*}
	\fontsize{8}{8}
	\begin{split}
		log_2(1 + g_f P^{IC})= log_2 \left (\frac{(1 + P_{a} g_f)(1 + (\alpha + \beta) P_{a} g_m)(1+ {\alpha P_{a} g_n})}{(1 + (\alpha + \beta) P_{a} g_f)(1+ {\alpha P_{a} g_m})}\right)
	\end{split}
\end{align*}
Let $P^{IC}$ = $P_{a}$ + $\zeta$, where -$p$ <  $\zeta < +p$, $p$ = $\big|P^{IC} - P_{a}\big|$, then from (\ref{18}), we have $\zeta$ > 0 $\implies$ $P^{IC}$ > $P_{a}$, i.e.,
\vspace{-0.2cm}
\begin{equation}
	\label{19}
	P_a = P^{IC} - \zeta
\end{equation}
Hence, a 3-Group NOMA transmission requires less power than the power needed in the conventional IC- scenario to achieve the same rate as the conventional IC system.

2)$\textit{CASE}$ II ($l_f>l_m>l_n$): For this case, we need to calculate the power per each 3-Group NOMA, 2-Group NOMA and IC transmission to ensure that the achievable information rate is at least as good as conventional IC.

From (\ref{19}), we know that the 3-Group NOMA transmissions in the  scheme require less power per transmission to support an information rate as good as conventional IC. Let $P_{(m,f)}^{(2)}$ denote the power per 2-Group NOMA transmission for this case. Then, (\ref{7}) is modified as

\vspace{-0.2cm}
\begin{equation}
	\small
	\begin{split}
		\label{20}
		R_{(m,f)}^{IN-NOMA^{(2)}} = log_2 \left (\frac{(1 +P_{(m,f)}^{(2)}g_f)(1 + \alpha_{1} P_{(m,f)}^{(2)} g_m)}{(1 + \alpha_{1}   P_{(m,f)}^{(2)} g_f)} \right).
	\end{split}
\end{equation}
To find the minimum power requirement  that makes  $R_{(m,f)}^{IN-NOMA^{(2)}}$ at least as good as $R^{IC}$, consider (\ref{16}) and (\ref{20}).
\vspace{-0.2cm}
\begin{equation*}
	\begin{split}
		log_2(1 + g_f P^{IC})=log_2 \left(\frac{(1 +P_{(m,f)}^{(2)}g_f)(1 + \alpha_{1} P_{(m,f)}^{(2)} g_m)}{(1 + \alpha_{1}   P_{(m,f)}^{(2)} g_f)}\right).
	\end{split}
\end{equation*}

Let $P^{IC}$ = $P_{(m,f)}^{(2)}$ + $\zeta_{(m,f)}$, where -$p_1 <  \zeta_{(m,f)} < +p_1$, $p_1$ = $\big|P^{IC} - P_{(m,f)}^{(2)}\big|$, then
\vspace{-0.2cm}
\begin{equation}
	\begin{split}
		\label{21}
		\zeta_{(m,f)} = \frac{(1+P_a g_f)(\alpha_{1} P_a (g_m - g_f))}{g_f(1 + \alpha_{1} P_a g_f)}.
	\end{split}
\end{equation}

From (\ref{21}), we have $\zeta_{(m,f)}$ > 0 $\implies$ $P^{IC}$ > $P_{(m,f)}^{(2)}$, i.e.,	\begin{equation}
	\label{22}
	P_{(m,f)}^{(2)}= P^{IC} - \zeta_{(m,f)}.
\end{equation}
Hence, a 2-Group NOMA transmission that combines index-coded packets meant for intermediate and far users requires less power than the power needed in the conventional IC- scenario to achieve the same rate as the conventional IC system.

Let $P_{f}^{IC}$ denote the power per IC transmission in Case-II. Since the IC transmissions in this case are meant for far group users, the power per IC transmission required in this case to achieve a rate at least as good as conventional IC will be equal to $P^{IC}$. Then, $P_{f}^{IC}=P^{IC}$.

3)$\textit{CASE}$ III ($l_n > l_f > l_m$) : The 3-Group NOMA transmission of the scheme saves power compared to conventional IC to achieve equal information rate as given by (\ref{19}).

Let $P_{(n,f)}^{(2)}$ denote the power per 2-Group NOMA transmission in this case. Then, (\ref{9}) is modified as 
\vspace{-0.2cm}
\begin{equation}
	\begin{split}
		\label{24}
		R_{(n,f)}^{IN-NOMA^{(2)}} = log_2 \left (\frac{(1 + P_{(n,f)}^{(2)} g_f)(1 + \alpha_{1} P_{(n,f)}^{(2)} g_n)}{(1 + \alpha_{1}   P_{(n,f)}^{(2)} g_f)} \right)
	\end{split}
\end{equation}
To find the minimum power requirement for a 2-Group IC-NOMA transmission that makes  $R_{(n,f)}^{IN-NOMA^{(2)}}$ at least as good as $R^{IC}$, consider (\ref{16}) and (\ref{24}).
\vspace{-0.2cm}
\begin{equation*}
	\small
	\begin{split}
		log_2(1 + g_f P^{IC})= log_2 \left (\frac{(1 + P_{(n,f)}^{(2)} g_f)(1 + \alpha_{1} P_{(n,f)}^{(2)} g_n)}{(1 + \alpha_{1}   P_{(n,f)}^{(2)} g_f)} \right)
	\end{split}
\end{equation*}

Let $P^{IC}$ = $P_{(n,f)}^{(2)}$ + $\zeta_{(n,f)}$, where -$p_2 <\zeta_{(n,f)} < +p_2$, \\$p_2$ = $\big|P^{IC} - P_{(n,f)}^{(2)}\big|$, then
\vspace{-0.2cm}
\begin{equation}
	\begin{split}
		\label{25}
		\zeta_{(n,f)} = \frac{(1+P_a g_f)(\alpha_{1} P_a (g_n - g_f))}{g_f(1 + \alpha_{1} P_a g_f)}.
	\end{split}
\end{equation}

From (\ref{25}), we have $\zeta_{(n,f)}$ > 0 $\implies$ $P^{IC}$ > $P_{(n,f)}^{(2)}$, i.e.,   
\vspace{-0.2cm}
\begin{equation}
	\label{26}
	P_{(n,f)}^{(2)}= P^{IC} - \zeta_{(n,f)}.
\end{equation}
Hence, a 2-Group NOMA transmission that combines index-coded packets meant for near and far users requires less power than the power needed in the conventional IC- scenario to achieve the same rate as the conventional IC system.

Let $P_{n}^{IC}$ denote the power per IC transmission in this case, then (\ref{1}) is modified as 
\vspace{-0.2cm}
\begin{equation}
	\label{27}
	R_n^{IN-IC} =  log_2(1+g_nP_{n}^{IC}).
\end{equation}

To find the minimum power per IC transmissions that makes $R_n^{IN-IC}$ at least as good as $R^{IC}$, consider (\ref{16}) and (\ref{27}). We have
\vspace{-0.2cm}
\begin{equation*}
	log_2(1+g_fP^{IC})=  log_2(1+g_nP_{n}^{IC}).
\end{equation*}

Let  $P^{IC}=P_{n}^{IC}+\zeta_{(n)}$, where  -$p_3 <  \zeta_{(n)} < +p_3$, $p_2$ = $\big|P^{IC} - P_{n}^{IC}\big|$, then
\vspace{-0.2cm}
\begin{equation}
	\label{28}
	\zeta_{(n)} =  \frac{(g_n - g_f)P_{n}^{IC}}{g_f}.
\end{equation}

From (\ref{28}), we have $\zeta_{(n)}$ > 0 $\implies$ $P^{IC} > P_{n}^{IC}$, i.e.,
\vspace{-0.2cm}
\begin{equation}
	\label{29}
	P_{n}^{IC}= P^{IC} - \zeta_{(n)}.
\end{equation}
		Hence, an IC transmission meant for the near user requires less power than the power needed in the conventional IC- scenario to achieve the same rate as the conventional IC system.

\begin{table*}
	\centering   				
	\caption{Four different cases of the 3-Group IC-NOMA system by chosing same known set and different Want set}
	\label{Tab:9}
	
	\begin{tabular}{|c|c|c|c|c|c|}
		\hline
		User & Known set & \multicolumn{4}{|c|}{Want set  \(\mathcal{W}\)$_i$ (After R2V Phase)}  \\
		\hline
		$V_i$ &  \(\mathcal{K}\)$_i$ & $l_f$ = $l_m$ = $l_n$ & $l_f$ > $l_m$ > $l_n$ & $l_n$ > $l_f$ > $l_m$ &$l_m$ > $l_n$ > $l_f$\\
		\hline
		$V_1$ & \{$x_{1},x_{2},x_{3}\}$ &  \{$x_{4},x_{5},x_{6},x_{8}\}$  &  \{$x_{4},x_{5},x_{6},x_{7}\}$  &  \{$x_{4},x_{5},x_{6}\}$  &  \{$x_{5},x_{6},x_{7},x_{9}\}$ \\
		\hline
		$V_2$ & \{$x_{2},x_{3},x_{4}\}$ &  \{$x_{1},x_{5},x_{6},x_{9}\}$  &  \{$x_{1},x_{5},x_{6},x_{8}\}$  &  \{$x_{1},x_{6},x_{7},x_{9}\}$  &  \{$x_{1},x_{4},x_{5},x_{7}\}$ \\
		\hline
		$V_3$ & \{$x_{3},x_{4},x_{5}\}$  &  \{$x_{1},x_{6},x_{7},x_{8}\}$  &  \{$x_{1},x_{2},x_{7},x_{9}\}$  &  \{$x_{2},x_{6},x_{7},x_{8}\}$  &  \{$x_{1},x_{6},x_{8}\}$ \\
		\hline
		$V_4$ & \{$x_{5},x_{6},x_{7}\}$ &  \{$x_{4},x_{8},x_{9}\}$  &  \{$x_{4},x_{8},x_{9}\}$  &  \{$x_{4},x_{8},x_{9}\}$  &  \{$x_{1},x_{2},x_{9}\}$ \\
		\hline
		$V_5$ & \{$x_{6},x_{7},x_{8}\}$ &  \{$x_{5},x_{8},x_{9}\}$  &  \{$x_{1},x_{9}\}$  &  \{$x_{4},x_{9}\}$  &  \{$x_{1},x_{5},x_{9}\}$ \\
		\hline
		$V_6$ & \{$x_{8},x_{9}\}$ &  \{$x_{7}\}$  &  \{$x_{6},x_{7}\}$  &  \{$x_{7}\}$  &  \{$x_{7}\}$ \\
		\hline
		$V_7$ & \{$x_{9}\}$ &  \{$x_{7},x_{8}\}$  &  \{$x_{6},x_{7},x_{8}\}$  &  \{$x_{7},x_{8}\}$  &  \{$x_{7}\}$ \\
		\hline						
	\end{tabular}
\end{table*}

\begin{table*}
	\centering   				
	\caption{Index codes designed for four different cases of 3-Group IC-NOMA system}
	\label{Tab:10}
	
	\begin{tabular}{|c|c|c|c|}
		\hline
		Case & \(\mathcal{Y}_f \) & \(\mathcal{Y}_m \) & \(\mathcal{Y}_n \) \\
		\hline	
		$l_f$ = $l_m$ = $l_n$  & 		 \{$x_{7}+x_{9},x_{8}+x_{9}\}$  & 		 \{$x_{4}+x_{8},x_{5}+x_{8}\}$ & 		 \{$x_{1}+x_{4},x_{3}+x_{6}\}$ \\
		\hline	
		$l_f$ > $l_m$ > $l_n$  & 		 \{$x_{6}+x_{9},x_{7}+x_{9}x_{8}+x_{9}\}$  & 		 \{$x_{1}+x_{7},x_{4}+x_{7}\}$ & 		 \{$x_{2}+x_{5}\}$ \\
		\hline	
		$l_n$ > $l_f$ > $l_m$ & 		 \{$x_{7}+x_{9},x_{8}+x_{9}\}$  & 		 \{$x_{4}+x_{8}\}$ & 		 \{$x_{1}+x_{4},x_{2}+x_{5},x_{3}+x_{6}\}$ \\
		\hline	
		$l_m$ > $l_n$ > $l_f$\  & 		 \{$x_{7}+x_{9}\}$  & 		 \{$x_{1}+x_{7},x_{2}+x_{9},x_{5}+x_{8}\}$ & 		 \{$x_{1}+x_{5},x_{3}+x_{6}\}$ \\
		\hline
	\end{tabular}
\end{table*}

4)$\textit{CASE}$ IV ($l_m > l_n > l_f$) : The 3-Group NOMA transmission of the  scheme saves power compared to conventional IC to achieve equal information rate as given by (\ref{19}).

Let $P_{(n,m)}^{(2)}$ denote the power per 2-Group NOMA transmission for Case-IV. Then (\ref{13}) is modified as 
\vspace{-0.2cm}
\begin{equation}       							
	\begin{split}
		\label{30}
		R_{(n,m)}^{IN-NOMA^{(2)}} = log_2 \left (\frac{(1 + P_{(n,m)}^{(2)} g_m)(1 + \alpha_{1} P_{(n,m)}^{(2)} g_n)}{(1 + \alpha_{1}   P_{(n,m)}^{(2)} g_m)} \right).
	\end{split}
\end{equation}
To find the minimum power requirement for 2-Group IC-NOMA transmission that makes  $R_{(n,m)}^{IN-NOMA^{(2)}}$ at least as good as $R^{IC}$, consider (\ref{16}) and (\ref{30}). We have
\vspace{-0.2cm}
\begin{equation*}
	\begin{split}
		log_2(1 + g_f P^{IC})=log_2 \left (\frac{(1 + P_{(n,m)}^{(2)} g_m)(1 + \alpha_{1} P_{(n,m)}^{(2)} g_n)}{(1 + \alpha_{1}   P_{(n,m)}^{(2)} g_m)} \right).
	\end{split}
\end{equation*}

Let $P^{IC}$ = $P_{(n,m)}^{(2)}$ + $\zeta_{(n,m)}$, where -$p_4 < \zeta_{(n,m)} < +p_4$,  \\
$p_4$ = $\big|P^{IC} - P_{(n,m)}^{(2)}\big|$, then
\vspace{-0.2cm}
\begin{equation}
	\small
	\begin{split}
		\label{31}
		\zeta_{(n,m)} = \frac{\alpha_{1} P_a (g_n-g_m) + P_a(g_m-g_f) + \alpha_{1} P_{a}^{2} g_m(g_n-g_f)}{g_f(1 + \alpha_{1} P_a g_f)}.
	\end{split}
\end{equation}

From (\ref{31}), we have $\zeta_{(n,m)}$ > 0 $\implies$ $P^{IC}$ > $P_{(n,m)}^{(2)}$, i.e.,
\vspace{-0.2cm}
\begin{equation}
	\label{32}
	P_{(n,m)}^{(2)}= P^{IC} - \zeta_{(n,m)}.
\end{equation}
Hence, a 2-Group NOMA transmission that combines index-coded packets meant for intermediate and near users requires less power than the power needed in the conventional IC- scenario to achieve the same rate as the conventional IC system.

Let $P_{m}^{IC}$ denote the power per IC transmission for Case-IV, then (\ref{2})  is modified as 
\vspace{-0.2cm}
\begin{equation}
	\label{33}
	R_m^{IN-IC} =  log_2(1+g_mP_{m}^{IC}).
\end{equation}

To find the minimum power per transmission for IC transmissions that makes $R_m^{IN-IC}$ at least as good as $R^{IC}$, consider (\ref{16}) and (\ref{33}) leading to 
\vspace{-0.2cm}
\begin{equation*}
	log_2(1+g_fP^{IC})= log_2(1+g_mP_{m}^{IC}).
\end{equation*}

Let  $P^{IC}$$=P_{m}^{IC}$$+\zeta_{(m)}$, where  -$p_5$ <  $\zeta_{(m)}$ < +$p_5$, $p_5$ = $\big|P^{IC} - P_{m}^{IC}\big|$, then
\vspace{-0.2cm}
\begin{equation}
	\label{34}
	\zeta_{(m)} =  \frac{(g_m - g_f)P_{m}^{IC}}{g_f}.
\end{equation}

From (\ref{34}), we have $\zeta_{(m)}$ > 0 $\implies$ $P^{IC} > P_{m}^{IC}$, i.e., 
\begin{equation}
	\label{35}
	P_{m}^{IC}= P^{IC} - \zeta_{(m)}.
\end{equation}
Hence, an IC transmission meant for the intermediate user requires less power than the power needed in the conventional IC- scenario to achieve the same rate as the conventional IC system.

The power savings for all the cases is given in  Table-\ref{Tab:8}. We can see that $P_{saving}^{(i)}$ $>$ 0, $\forall$ $i$ $\in$ [13].

\section{ILLUSTRATION OF RATE AND TRANSMISSION POWER ANALYSIS}
\label{Illustarions}
In this section we illustrate our transmission scheme and index code construction for an example and also show the improvement in the performance of the 3-Group IC-NOMA system when compared with the conventional index coding system. The example considered is a VANET scenario with $N=7$ users having the known set and want set conditions as given in Table- \ref{Tab:9}. We classify users $V_1, V_2, V_3$ as near group users, users $V_4, V_5$ as intermediate group users and $V_6, V_7$ as far group users.
Table-\ref{Tab:10} shows the index code designed for four want set conditions developed using Algorithm-\ref{alg:1}.

\subsection{\textit{Rate Analysis}}
	Fig. \ref{figure1} compares the average achievable rate of the proposed scheme with the conventional index coding scenario. In each case of the 3-Group IC-NOMA system, a total of  $l^{IC-NOMA}$  transmissions will be there with $l^{NOMA^{(3)}}$ 3-Group NOMA transmissions, $l^{NOMA^{(2)}}$ 2-Group NOMA transmissions and $l^{IN-IC}$ IC transmissions. The average information rate, $R_{avg}$ for each case of the 3-Group IC-NOMA system is calculated as

\begin{figure}
			\centering
			\includegraphics[width= \columnwidth]{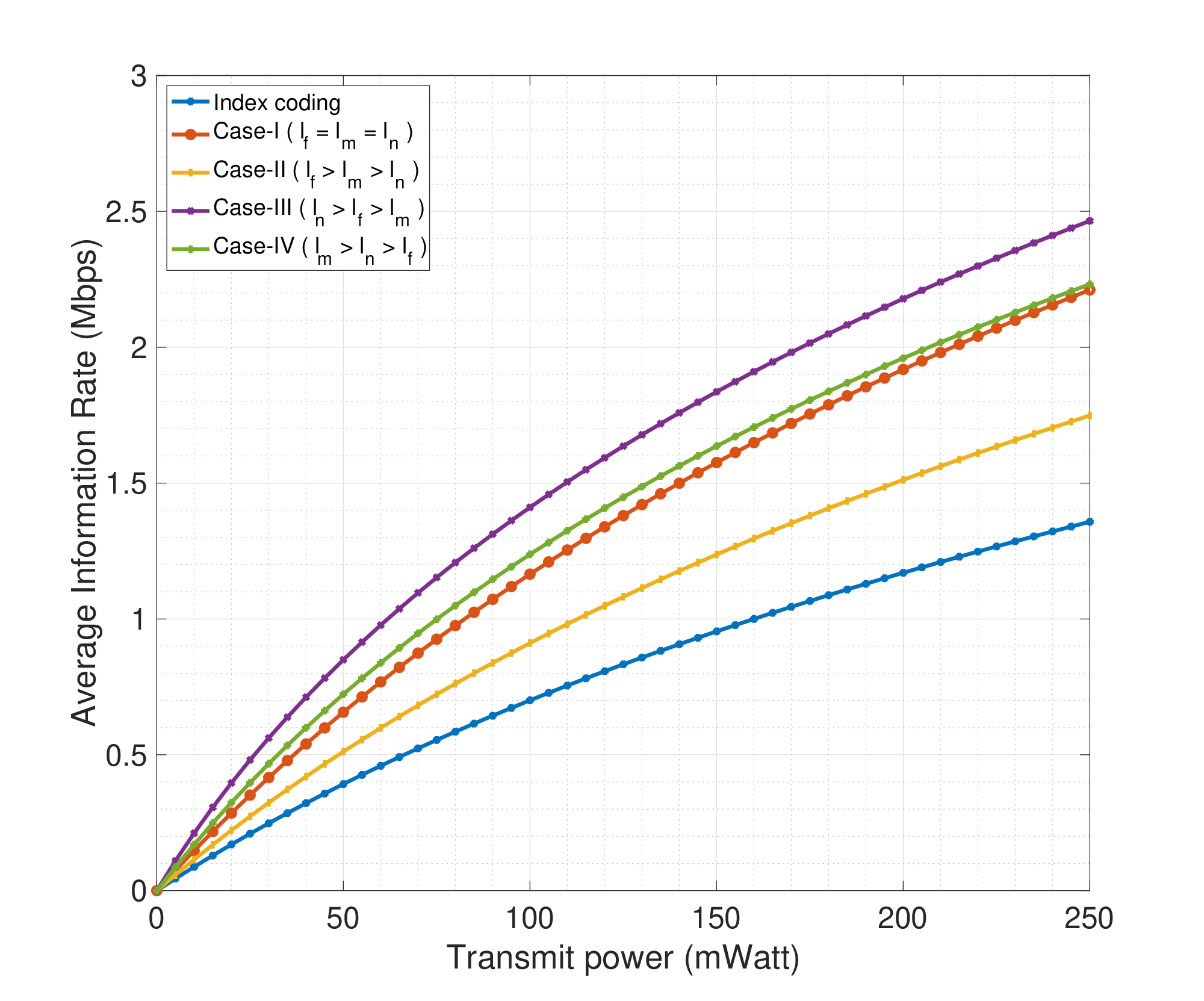}
	\caption{Comparison of the average information rate for the four cases of the proposed system as given in Table-\ref{Tab:9} with that of conventional IC.}
	\label{figure1}
\end{figure}
\vspace{-0.6cm}
\begin{equation}
	R_{avg}= \frac{Sum \hspace{0.1cm} of \hspace{0.1cm} rates \hspace{0.1cm} of \hspace{0.1cm} all\hspace{0.1cm} transmissions}{l^{IC-NOMA}}.
\end{equation}
In the considered cases, the plot shows that the average information rate is higher for Case III than for others. This is because the information rate for IC transmissions, in this case, is determined by the channel gains of near users. The average information rate is the least for Case II as the channel gains of far users determine the information rate for IC and 2-Group NOMA transmissions in Case II.

\subsection{\textit{Transmission Power Analysis}}
							Fig. \ref{fig2} compares the average power required to achieve the same information rate for conventional IC and 3-Group IC-NOMA systems. Each case's average power consumption $P_{avg}$ is calculated on similar lines as $R_{avg}$. Here, it can be seen that the 3-Group IC-NOMA system saves power compared to conventional IC for achieving the same information rate. Similar to the results of Fig.\ref{figure1}, the power saving is also highest for Case III and least for Case II.

\begin{figure}
	\centering
	\includegraphics[width= \columnwidth]{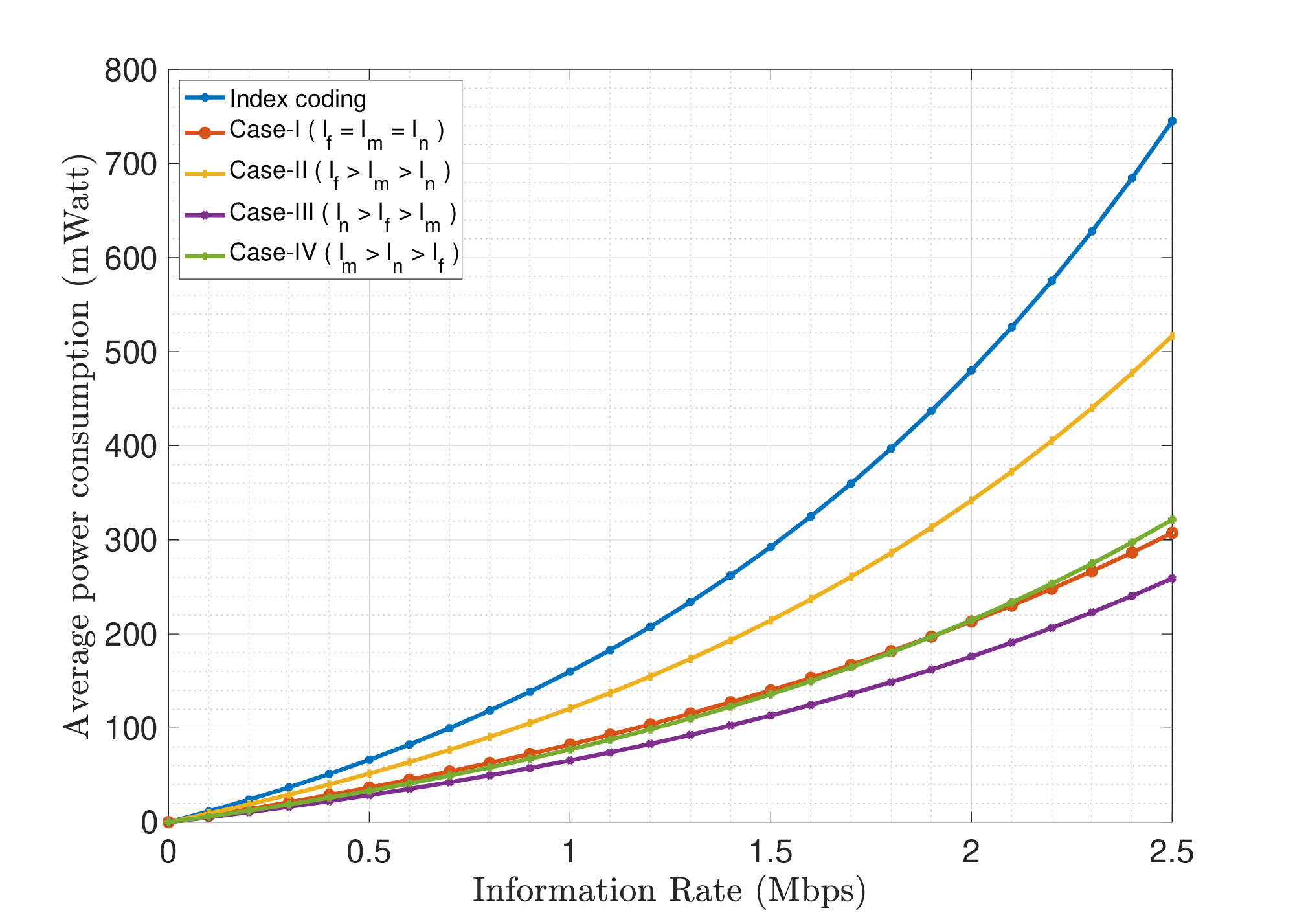}
	\caption{Comparison of the average power consumption for the four cases of the proposed system as given in Table-\ref{Tab:9} with that of conventional IC.}
		\label{fig2}
\end{figure}

\section{Appendix-1} 
\label{APP1}
\begin{center}
RATE ANALYSIS FOR CASES V to XIII
\end{center}

5) $\textit{CASE}$ V ($l_f>l_m=l_n$) : In this case, a total of $l_f$ transmissions will be required by 3-Group IC-NOMA scheme, with $l_n$  3-Group NOMA transmissions and $l_f-l_m$ index coded transmissions. The transmitted messages are

\begin{equation*}
	\fontsize{8}{8}
	S_{k}^{IC-NOMA} = \begin{cases}
		
		\sqrt{\alpha P}s_{n_{k}} +\sqrt{\beta P}s_{m_{k}}+\sqrt{\gamma P}s_{f_{k}}, k \in [l_n] \\[1ex]       		
		and\\
		\sqrt{P}s_{f_{k}}                 ,  k=\{l_n+1,\cdots l_f\}\\[1ex]
	\end{cases}
\end{equation*}

with $\alpha<\beta<\gamma$  and $\alpha+\beta+\gamma$=1  

The information rate for 3-Group NOMA transmissions will be the same as (\ref{5}).
In this case, the index-coded transmissions are index-coded packets designed for far users. Hence, 
the achievable rate for IC transmissions is given by (\ref{3}).

Therefore, for this case, the improvement in the information rate supported by 3-Group NOMA transmissions is given by (\ref{6}). In this case, the achievable rate of IC transmissions is the same as that of conventional IC.

6) $\textit{CASE}$ VI ( $l_f > l_n > l_m$) : In this case, a total of $l_f$ transmissions will be required by 3-Group IC-NOMA scheme, with $l_m$  3-Group NOMA transmissions, $l_n-l_m$ 2-Group NOMA transmissions and $l_f-l_n$ index-coded transmissions. The transmitted messages are

\begin{equation*}
	\fontsize{8}{8}
	S_{k}^{IC-NOMA} = \begin{cases}
		
		\sqrt{\alpha P}s_{n_{k}} +\sqrt{\beta P}s_{m_{k}}+\sqrt{\gamma P}s_{f_{k}}, k \in [l_m] \\[1ex]
		
		\sqrt{\alpha_{1} P}s_{n_{k}}+\sqrt{(1-\alpha_{1})P}s_{f_{k}},k=\{l_m+1,\cdots l_n\}\\[1ex]
		and\\
		\sqrt{P}s_{f_{k}}                 ,  k=\{l_n+1,\cdots l_f\}\\[1ex]
	\end{cases}
\end{equation*}

with $\alpha<\beta<\gamma$, $\alpha+\beta+\gamma$=1  and $\alpha_{1}$ < 0.5 

The information rate for 3-Group NOMA transmissions will be the same as (\ref{5}). The information rate for 2-Group NOMA transmissions will be the same as (\ref{9}). In this case, the index-coded transmissions are index-coded packets designed for far users. Hence, 
the achievable rate for IC transmissions is given by (\ref{3}).

Therefore, for this case, the improvement in the information rate supported by 3-Group NOMA transmissions and 2-Group NOMA transmissions is given by (\ref{6}) and (\ref{10}), respectively. In this case, the achievable rate of IC transmissions is the same as that of conventional IC.

7) $\textit{CASE}$ VII ($l_f = l_n > l_m$) : In this case, a total of $l_f$ transmissions will be required by 3-Group IC-NOMA scheme, with $l_m$  3-Group NOMA transmissions and $l_n-l_m$ 2-Group NOMA transmissions.

\begin{equation*}
	\fontsize{8}{8}
	S_{k}^{IC-NOMA} = \begin{cases}
		
		\sqrt{\alpha P}s_{n_{k}} +\sqrt{\beta P}s_{m_{k}}+\sqrt{\gamma P}s_{f_{k}}, k \in [l_m] \\[1ex]
		and\\
		\sqrt{\alpha_{1} P}s_{n_{k}}+\sqrt{(1-\alpha_{1})P}s_{f_{k}},k=\{l_m+1,\cdots l_f\}\\[1ex]
		
	\end{cases}
\end{equation*}

with $\alpha<\beta<\gamma$, $\alpha+\beta+\gamma$=1  and $\alpha_{1}$ < 0.5 

The information rate for 3-Group NOMA transmissions will be the same as (\ref{5}). The information rate for 2-Group NOMA transmissions will be the same as  (\ref{9}).

Therefore, for this case, the improvement in the information rate supported by 3-Group NOMA transmissions and 2-Group NOMA transmissions is given by (\ref{6}) and (\ref{10}), respectively.

8) $\textit{CASE}$ VIII ($l_f = l_m > l_n$) : In this case, a total of $l_f$ transmissions will be required by 3-Group IC-NOMA scheme, with $l_n$  3-Group NOMA transmissions  and $l_m-l_n$ 2-Group NOMA transmissions . The transmitted messages are

\begin{equation*}
	\fontsize{8}{8}
	S_{k}^{IC-NOMA} = \begin{cases}
		
		\sqrt{\alpha P}s_{n_{k}} +\sqrt{\beta P}s_{m_{k}}+\sqrt{\gamma P}s_{f_{k}}, k \in [l_n] \\[1ex]
		and\\
		\sqrt{\alpha_{1} P}s_{m_{k}}+\sqrt{(1-\alpha_{1})P}s_{f_{k}},k=\{l_n+1,\cdots l_f\}\\[1ex]

	\end{cases}
\end{equation*}

with $\alpha<\beta<\gamma$, $\alpha+\beta+\gamma$=1  and $\alpha_{1}$ < 0.5 

The information rate for 3-Group NOMA transmissions will be the same as (\ref{5}). The information rate for 2-Group NOMA transmissions will be the same as (\ref{7}).

Therefore, for this case, the improvement in the information rate supported by 3-Group NOMA transmissions and 2-Group NOMA transmissions is given by (\ref{6}) and (\ref{8}), respectively.

9) $\textit{CASE}$ IX ($l_n > l_f = l_m$) : In this case, a total of $l_n$ transmissions will be required by 3-Group IC-NOMA scheme, with $l_m$  3-Group NOMA transmissions and $l_n-l_f$ index-coded transmissions. The transmitted messages are

\begin{equation*}
	\fontsize{8}{8}
	S_{k}^{IC-NOMA} = \begin{cases}
		
		\sqrt{\alpha P}s_{n_{k}} +\sqrt{\beta P}s_{m_{k}}+\sqrt{\gamma P}s_{f_{k}}, k \in [l_m] \\[1ex]         		
		and\\
		\sqrt{P}s_{n_{k}}                 ,  k=\{l_m+1,\cdots l_n\}\\[1ex]
	\end{cases}
\end{equation*}

with $\alpha<\beta<\gamma$ and $\alpha+\beta+\gamma$=1  

The information rate for 3-Group NOMA transmissions will be the same as (\ref{5}).  In this case, the index-coded transmissions are index-coded packets designed for near users. Hence, 
the achievable rate for IC transmissions is given by (\ref{1}).

Therefore, for this case, the improvement in the information rate supported by 3-Group NOMA transmissions is given by (\ref{6}). From (\ref{11}), it can be seen that the achievable rate of IC transmissions is greater when compared to conventional IC.

10) $\textit{CASE}$ X ($l_n = l_m > l_f$) : In this case, a total of $l_n$ transmissions will be required by 3-Group IC-NOMA scheme, with $l_f$  3-Group NOMA transmissions  and $l_m-l_f$ 2-Group NOMA transmissions. The transmitted messages are

\begin{equation*}
	\fontsize{8}{8}
	S_{k}^{IC-NOMA} = \begin{cases}
		
		\sqrt{\alpha P}s_{n_{k}} +\sqrt{\beta P}s_{m_{k}}+\sqrt{\gamma P}s_{f_{k}}, k \in [l_f] \\[1ex]
		and\\
		\sqrt{\alpha_{1} P}s_{n_{k}}+\sqrt{(1-\alpha_{1})P}s_{m_{k}},k=\{l_f+1,\cdots l_n\}\\[1ex]
	\end{cases}
\end{equation*}

with $\alpha<\beta<\gamma$, $\alpha+\beta+\gamma$=1  and $\alpha_{1}$ < 0.5 

The information rate for 3-Group NOMA transmissions will be the same as (\ref{5}). The information rate for 2-Group NOMA transmissions will be the same as (\ref{13}).

Therefore, for this case, the improvement in the information rate supported by 3-Group NOMA transmissions and 2-Group NOMA transmissions is given by (\ref{6}) and (\ref{14}), respectively.

11) $\textit{CASE}$ XI ( $l_m > l_f = l_n$) : In this case, a total of $l_m$ transmissions will be required by 3-Group IC-NOMA scheme, with $l_n$  3-Group NOMA transmissions and $l_m-l_f$ index-coded transmissions. The transmitted messages are

\begin{equation*}
	\fontsize{8}{8}
	S_{k}^{IC-NOMA} = \begin{cases}
		
		\sqrt{\alpha P}s_{n_{k}} +\sqrt{\beta P}s_{m_{k}}+\sqrt{\gamma P}s_{f_{k}}, k \in [l_n] \\[1ex]
		and\\
		\sqrt{P}s_{m_{k}}                 ,  k=\{l_n+1,\cdots l_m\}\\[1ex]
		
	\end{cases}
\end{equation*}

with $\alpha<\beta<\gamma$  and $\alpha+\beta+\gamma$=1.

The information rate for 3-Group NOMA transmissions will be the same as (\ref{5}).  In this case, the index-coded transmissions are index-coded packets designed for intermediate users. Hence, 
the achievable rate for IC transmissions is given by (\ref{2}).

Therefore, for this case, the improvement in the information rate supported by 3-Group NOMA transmissions is given by (\ref{6}). From (\ref{15}), it can be seen that the achievable rate of IC transmissions is greater when compared to conventional IC.

12) $\textit{CASE}$ XII ($l_n > l_m > l_f$) : In this case, a total of $l_n$ transmissions will be required by 3-Group IC-NOMA scheme, with $l_f$  3-Group NOMA transmissions, $l_m-l_f$ 2-Group NOMA transmissions and $l_n-l_m$ index-coded transmissions. The transmitted messages are

\begin{equation*}
	\fontsize{8}{8}
	S_{k}^{IC-NOMA} = \begin{cases}
		
		\sqrt{\alpha P}s_{n_{k}} +\sqrt{\beta P}s_{m_{k}}+\sqrt{\gamma P}s_{f_{k}}, k \in [l_f] \\[1ex]
		
		\sqrt{\alpha_{1} P}s_{n_{k}}+\sqrt{(1-\alpha_{1})P}s_{m_{k}},k=\{l_f+1,\cdots l_m\}\\[1ex]
		and\\
		\sqrt{P}s_{n_{k}}                 ,  k=\{l_m+1,\cdots l_n\}\\[1ex]
	\end{cases}
\end{equation*}

with $\alpha<\beta<\gamma$, $\alpha+\beta+\gamma$=1  and $\alpha_{1}$ < 0.5 

The information rate for 3-Group NOMA transmissions will be the same as (\ref{5}). The information rate for 2-Group NOMA transmissions will be the same as (\ref{13}). In this case, the index-coded transmissions are index-coded packets designed for near users. Hence, 
the achievable rate for IC transmissions is given by (\ref{1}).

Therefore, for this case, the improvement in the information rate supported by 3-Group NOMA transmissions and 2-Group NOMA transmissions is given by (\ref{6}) and (\ref{14}), respectively. From (\ref{11}), it can be seen that the achievable rate of IC transmissions is greater when compared to conventional IC.

13) $\textit{CASE}$ XIII ($l_m > l_f > l_n$) : In this case, a total of $l_m$ transmissions will be required by 3-Group IC-NOMA scheme, with $l_n$  3-Group NOMA transmissions, $l_f-l_n$ 2-Group NOMA transmissions and $l_m-l_f$ index-coded transmissions. The transmitted messages are

\begin{equation*}
	\fontsize{8}{8}
	S_{k}^{IC-NOMA} = \begin{cases}
		
		\sqrt{\alpha P}s_{n_{k}} +\sqrt{\beta P}s_{m_{k}}+\sqrt{\gamma P}s_{f_{k}}, k \in [l_n] \\[1ex]
		
		\sqrt{\alpha_{1} P}s_{m_{k}}+\sqrt{(1-\alpha_{1})P}s_{f_{k}},k=\{l_n+1,\cdots l_f\}\\[1ex]
		and\\
		\sqrt{P}s_{m_{k}}                 ,  k=\{l_f+1,\cdots l_m\}\\[1ex]
	\end{cases}
\end{equation*}

with $\alpha<\beta<\gamma$, $\alpha+\beta+\gamma$=1  and $\alpha_{1}$ < 0.5 

The information rate for 3-Group NOMA transmissions will be the same as (\ref{5}). The information rate for 2-Group NOMA transmissions will be the same as (\ref{7}). In this case, the index-coded transmissions are index-coded packets designed for intermediate users. Hence, 
the achievable rate for IC transmissions is given by (\ref{2}).

Therefore, for this case, the improvement in the information rate supported by 3-Group NOMA transmissions and 2-Group NOMA transmissions is given by (\ref{6}) and (\ref{8}), respectively. From (\ref{15}), it can be seen that the achievable rate of IC transmissions is greater when compared to conventional IC.

\section{Appendix-2}
\label{APP2}
\begin{center}
	TRANSMISSION POWER ANALYSIS FOR CASES V to XIII
\end{center}
5) $\textit{CASE}$ V ($l_f>l_m=l_n$): The 3-Group NOMA transmission of the 3-Group IC-NOMA scheme saves power compared to conventional IC to achieve equal information rate as given by (\ref{19}). We have zero power savings for IC transmissions in this case. The power savings, in this case, is given by
\begin{align*}
	P_{saving}^{(5)} =
	\nonumber & (l^{IC}-l_{n})P^{IC} + \zeta l_n 
\end{align*}

6) $\textit{CASE}$ VI ( $l_f > l_n > l_m$): The 3-Group NOMA transmission of the 3-Group IC-NOMA scheme saves power compared to conventional IC to achieve equal information rate as given by (\ref{19}).  For this case, the 2-group NOMA transmission of the 3-Group IC-NOMA scheme saves power compared to conventional IC to achieve an equal information rate is given by (\ref{26}). We have zero power savings for IC transmissions in this case. The power savings, in this case, is given by
\begin{align*}
	P_{saving}^{(6)} =
	\nonumber & (l^{IC}-l_{n})P^{IC} + \zeta l_m  + (l_n-l_m)\zeta_{(n,f)} 
\end{align*}

7) $\textit{CASE}$ VII ($l_f = l_n > l_m$): The 3-Group NOMA transmission of the 3-Group IC-NOMA scheme saves power compared to conventional IC to achieve equal information rate as given by (\ref{19}).  For this case, the 2-group NOMA transmission of the 3-Group IC-NOMA scheme saves power compared to conventional IC to achieve an equal information rate is given by (\ref{26}). The power savings, in this case, is given by
\begin{align*}
	P_{saving}^{(7)} =
	\nonumber & (l^{IC}-l_{n})P^{IC} + \zeta l_m  + (l_n-l_m)\zeta_{(n,f)}
\end{align*}

8) $\textit{CASE}$ VIII ($l_f = l_m > l_n$): The 3-Group NOMA transmission of the 3-Group IC-NOMA scheme saves power compared to conventional IC to achieve equal information rate as given by (\ref{19}). For this case, the 2-group NOMA transmission of the 3-Group IC-NOMA scheme saves power compared to conventional IC to achieve an equal information rate is given by (\ref{22}). The power savings, in this case, is given by
\begin{align*}
	P_{saving}^{(8)} =
	\nonumber & (l^{IC}-l_{m})P^{IC} + \zeta l_n  + (l_m-l_n)\zeta_{(m,f)} 
\end{align*}

9) $\textit{CASE}$ IX ($l_n > l_f = l_m$): The 3-Group NOMA transmission of the 3-Group IC-NOMA scheme saves power compared to conventional IC to achieve equal information rate as given by (\ref{19}). In this case, we have non-zero power savings for IC transmissions as given by (\ref{29}). The power savings, in this case, is given by
\begin{align*}
	P_{saving}^{(9)} =
	\nonumber & (l^{IC}-l_{n})P^{IC} + \zeta l_m  +  (l_n-l_f)\zeta_{(n)}
\end{align*}

10) $\textit{CASE}$ X ($l_n = l_m > l_f$): The 3-Group NOMA transmission of the 3-Group IC-NOMA scheme saves power compared to conventional IC to achieve equal information rate as given by (\ref{19}).  For this case, the 2-group NOMA transmission of the 3-Group IC-NOMA scheme saves power compared to conventional IC to achieve an equal information rate is given by (\ref{32}). The power savings, in this case, is given by
\begin{align*}
	P_{saving}^{(10)} =
	\nonumber & (l^{IC}-l_{m})P^{IC} + \zeta l_f  + (l_m-l_f)\zeta_{(n,m)} 
\end{align*}

11) $\textit{CASE}$ XI ( $l_m > l_f = l_n$): The 3-Group NOMA transmission of the 3-Group IC-NOMA scheme saves power compared to conventional IC to achieve equal information rate as given by (\ref{19}). In this case, we have non-zero power savings for IC transmissions as given by  (\ref{35}). The power savings, in this case, is given by
\begin{align*}
	P_{saving}^{(11)} =
	\nonumber & (l^{IC}-l_{m})P^{IC} + \zeta l_n  + (l_m-l_f)\zeta_{(m)}
\end{align*}

12) $\textit{CASE}$ XII ($l_n > l_m > l_f$): The 3-Group NOMA transmission of the 3-Group IC-NOMA scheme saves power compared to conventional IC to achieve equal information rate as given by (\ref{19}). For this case, the 2-group NOMA transmission of the 3-Group IC-NOMA scheme saves power compared to conventional IC to achieve an equal information rate is given by (\ref{32}). In this case, we have non-zero power savings for IC transmissions as given by (\ref{29}). The power savings, in this case, is given by
\begin{align*}
	P_{saving}^{(12)} =
	\nonumber &(l^{IC}-l_{n})P^{IC} + \zeta l_f  + (l_m-l_f)\zeta_{(n,m)} + \\&(l_n-l_m)\zeta_{(n)}
\end{align*}

13) $\textit{CASE}$ XIII ($l_m > l_f > l_n$): The 3-Group NOMA transmission of the 3-Group IC-NOMA scheme saves power compared to conventional IC to achieve equal information rate as given by (\ref{19}). For this case, the 2-group NOMA transmission of the 3-Group IC-NOMA scheme saves power compared to conventional IC to achieve an equal information rate is given by (\ref{22}). We have non-zero power savings for IC transmissions, as given by (\ref{35}). the power savings, in this case, is given by
\begin{align*}
	P_{saving}^{(13)} =
	\nonumber &(l^{IC}-l_{m})P^{IC} + \zeta l_n  + (l_f-l_n)\zeta_{(m,f)} + \\& (l_m-l_f)\zeta_{(m)}
\end{align*}

\section{Conclusion}
This paper proposes a spectral efficient transmission scheme called 3-Group IC-NOMA for VANETs and presents an algorithm to design index codes for the proposed scheme, which allows to reap the benefits of NOMA. We have demonstrated the construction of different transmissions for the 3-Group IC-NOMA transmission scheme using Algorithm 1  and proved that the proposed system provided non-negative bandwidth savings compared to conventional IC systems by serving user demands with less than or an equal number of transmissions compared to traditional IC. Based on detailed analytical studies, we could conclude that this transmission scheme supports higher information rates than traditional IC systems. We have also proved that the proposed transmission scheme always has positive power savings to achieve a rate at least as good as conventional IC.

\section{Acknowledgment}
This work was supported partly by the Science and Engineering Research Board (SERB) of Department of Science and	Technology (DST), Government of India, through J.C. Bose National Fellowship to B. Sundar Rajan.

\end{document}